\documentclass[twocolumn,aps,prb,preprintnumbers,footinbib,floatfix,superscriptaddress]{revtex4-2}
\usepackage{amsmath}
\usepackage{amssymb}
\usepackage{color}
\usepackage{graphicx}

\usepackage{amsmath,amsfonts,amsthm}
\usepackage{graphicx}
\usepackage{float} \usepackage{subfig}
\usepackage{booktabs}
\usepackage{lineno}

\begin{document}

\title{Pyroresistive response of percolating conductive polymer composites}

\author{Ettore Barbieri}
\affiliation{Center for Mathematical Science and Advanced Technology (MAT), Research Institute for Value-Added-Information Generation (VAiG), Japan Agency for Marine-Earth Science and Technology, Yokohama Institute for Earth Sciences, 3173-25, Showa-machi, Kanazawa-ku, Yokohama City, Kanagawa 236-0001, Japan}
\author{Emiliano Bilotti}\affiliation{Department of Aeronautics, Imperial College London, South Kensington, London SW7 2AZ, UK}
\author{Yi Liu}\affiliation{Department of Materials, Loughborough University, Loughborough, LE11 3TU, UK}
\author{Claudio Grimaldi}\affiliation{Laboratory of Statistical Biophysics, Ecole Polytechnique F\'ed\'erale de
Lausanne, Station 3, CH-1015 Lausanne, Switzerland}
\affiliation{Centro Studi e Ricerche Enrico Fermi, I-00184 Roma, Italy}

\begin{abstract}
The pyroresistive response of conductive polymer composites (CPCs) has attracted much interest because of its potential applications in
many electronic devices requiring a significant responsiveness to changes in external physical parameters such as temperature or electric fields.
Although extensive research has been conducted to study how the properties of the polymeric matrix and conductive fillers
affect the positive temperature coefficient pyroresistive effect, the understanding of the microscopic mechanism governing
such a phenomenon is still incomplete.
In particular, to date, there is little body of theoretical research devoted to investigating the effect of the polymer
thermal expansion on the electrical connectivity of the conductive phase. 
Here, we present the results of simulations of model CPCs in which rigid conductive fillers are dispersed in an insulating 
amorphous matrix. By employing a meshless algorithm to analyze the thermoelastic response of the system, we couple the
computed strain field to the electrical connectedness of the percolating conductive particles. 
We show that the electrical conductivity responds to the local strains that are generated by the
mismatch between the thermal expansion of the polymeric and conductive phases and that the conductor-insulator transition is 
caused by a sudden and global disconnection of the electrical contacts forming the percolating network.

\end{abstract}

\maketitle

\section{introduction}
\label{intro}

Functional materials exhibiting novel combinations of mechanical, thermal, and electrical properties present great interest for various technological 
applications that require electrical materials that are lightweight, flexible, wearable, or that respond in a controlled manner to external stimuli. In
this context, conductive polymer composites (CPCs) fulfill several of these requirements and have been exploited in a wide range of technological applications.

In this class of composites, the electrical conductivity is established by the incorporation of conductive fillers into the otherwise insulating polymer 
matrix, forming a network of electrically connected conductive particles spanning the entire system \cite{Zhang2007}.
Typical examples of CPCs are polymers blended with carbon black particles, carbon fibers, carbon nanotubes and graphene-related materials, 
but also with metallic fillers such as Ag or Ni particles. By tuning the amount, size, dispersion and shape of the conductive fillers, CPCs can be tailored to combine advantageous properties of the polymer matrix (e.g., flexibility, toughness, processability, low density, etc.) with appropriate levels of electrical conductivity.

Among the several properties that are of interest for industrial application, a range of CPCs also exhibit a pronounced pyroresistive effect, 
manifested by a sharp increase in electrical resistivity with increasing temperature - a phenomenon known as the positive temperature coefficient 
(PTC) effect - which is exploited in current limiting devices and thermal switchers \cite{Xu2016,Liu2019,Liu2020,Liu2023}.

The PTC effect is generally understood as being due to the mismatch between the temperature-induced volume change of the polymer and the conducting phase. 
As the temperature increases, the larger thermal expansion of the polymer 
compared to that of the conducting particles entails a greater separation between the fillers, thus weakening or even breaking the macroscopic conducting 
path through the composite. This results in a sharp increase in electrical resistivity upon heating, often converting the CPC material from an electrical conductor 
to an insulator. 

Typically, the onset (or switching) temperature of the PTC effect (hereafter, $T_0$)  
corresponds to the temperature $T_m$ at which the polymer experiences a phase transition (e.g., melting of the crystalline phase,
a glass transition, or a rubber-liquid transition) accompanied by a sudden expansion of the polymeric matrix \cite{Luo2000,Li2002,Shen2007,Kar2010,Kar2012}.
In a few cases, though, significantly lower values of $T_0$ compared to $T_m$ have been documented within a specific range 
of filler content and as a function of the content itself. Such unusual results have been primarily observed in CPCs charged with
metal particles, typically of micrometric size, at relatively high loadings, and when subjected to very low heating rates \cite{Kar2010,Kar2012,Xi2004,Mitsuhiro2005,Rybak2010,Kono2012,Zhang2014,Xu2022}. 

It is not surprising that the particle size plays a role in the pyroresistive response, as it generally influences the overall behavior
of the electrical transport in composites. In addition to the higher specific surface area of small particles that may affect the conductive phase's dispersion within the polymer, the tunneling 
decay length $\xi$, which ranges from a fraction to a few nanometers, represents a discriminating factor. 
Conductive particles of dimensions much larger than $\xi$ can be considered electrically connected only if they
essentially touch each other, so the electrical conductivity of the composite is non-zero only if there exists a percolating 
cluster of particles at contact that spans the entire composite \cite{Balberg2009}. In this case, even a small increase in the particles separation
would entail a significant enhancement of the interparticle resistance and eventually the disruption of the conducting network. On the contrary,
in composites with nanometric fillers, the interparticle tunneling processes may extend beyond nearest neighbours \cite{Ambrosetti2010b}, thereby increasing  
the number of conductive pathways and making the conductive network more resilient to small increases in particle separation.
In such instances, the PTC effect is expected only if the polymer undergoes a significant volume expansion, induced, for example,
by a structural transition.

Besides the size of fillers used in CPCs, other factors influence the pyroresistive response. These include, among others, 
the shape and aspect ratio of the conducting fillers \cite{Asare2016}, their thermal conductance, the degree of polymer crystallization, or the type of dispersion of the conductive particles inside the 
polymer matrix. The combination of these factors and the incomplete understanding of their relative importance in influencing the 
pyroresistive response preclude a fully quantitative explanation of the PTC effect and how to systematically control it \cite{Xu2011}.

In particular, on the theoretical side, there is a lack of in-depth studies exploring the microscopic mechanism of pyroresistivity and 
how the interparticle connectivity evolves in response to uneven strain fields induced by the mismatch between the thermal expansion 
of the polymer and the conducting phase. This is especially important for
CPCs with large conductive fillers (of the order of a micrometer or more) since, as mentioned above, in this case, the electrical
connectivity between the particles is expected to be particularly sensitive to the local thermal expansion of the polymer matrix.

Here we explore the pyroresistive response of a continuous model of percolating CPCs by considering the coupling between local thermal 
strains of the polymer matrix and the electrical connectivity of rigid spherical conductive particles. By numerically computing 
the strain field within the continuous matrix, we calculate the strain-induced displacement of the conductive spheres to obtain the pyroresistive
response for different volume fractions, thermal strains and dispersions of the conductive phase.

\section{Model and simulations}
\label{model}

In general, conductor-insulator composites with conductive fillers with dimensions of the order of a micrometer or more
can be treated as \textit{bona fide} percolating systems, 
in which two neighboring conductive particles can be considered either electrically connected or disconnected depending on whether 
they are physically in contact or not. This must be contrasted to the electrical connectivity between conducting particles with
nanometric dimensions, in which even for particles that do not physically touch each other, their mean separation can be such that 
electrons can still flow from one particle to the others via tunneling processes.

As mentioned in the introduction, percolating composites typically exhibit a greater sensitivity of the electrical connectivity
to the thermal expansion of the polymeric phase compared to those primarily governed by tunneling, assuming similar conductive 
particle loadings. Hence, we focus our analysis on percolating CPCs (that is, CPCs charged with micrometric conductive fillers)
as these enable the calculation of the pyroresistive effect even with minor volume expansions, where linear elasticity remains applicable.

A direct consequence of assuming an on-off mechanism of electrical connectivity between the fillers is that the electrical conductivity of the composite
is non-zero only if there exists a macroscopic cluster of connected particles that spans the entire composite \cite{Balberg2009,Ambrosetti2010b}. This is reflected by a sharp increase of
the bulk conductivity $\sigma$ when the amount of the conducting phase increases beyond a critical value $\phi_c$, commonly referred to as the percolation
threshold \cite{Stauffer1994}. For values of the volume fraction $\phi$ close but above $\phi_c$, the conductivity follows, in this case, a power-law of the form \cite{Stauffer1994,Sahimi2003}: 
\begin{equation}
\label{sigma1}
\sigma\propto (\phi-\phi_c)^t,
\end{equation}
where $t$ is a universal critical exponent taking the value $t\simeq 2.0$ for all three-dimensional percolating systems, regardless of the microscopic details \cite{Sahimi2003}.

Contrary to the exponent $t$, the percolation threshold value strongly depends on the shape of the conductive fillers and their 
dispersion within the insulating matrix.  CPCs prepared with high aspect-ratios fillers, such as carbon fibers and carbon nanotubes, 
have systematically smaller percolation thresholds than CPCs with spherically shaped fillers \cite{Kyrylyuk2008,Bauhofer2009,Mutiso2015}. The latter, however, can display comparatively small values of $\phi_c$ if the 
conductive phase is not homogeneously dispersed in the matrix but is instead segregated within a smaller region of space, for instance
the amorphous phase of semi-crystalline polymers or one of the phases of immiscible polymer blends \cite{Pang2014}. 

In the following, we focus in particular on the difference in the pyroresistive response between model systems of homogeneous and segregated 
dispersions of spherical conductive particles in an amorphous (polymer) matrix.

\subsection{Contact algorithm}
\label{contact}
In formulating a model of percolating CPCs which can display a range of $\phi_c$ values,  we construct distributions of impenetrable and identical 
spherical conductive particles of diameter $D$  that are randomly dispersed with different degrees of heterogeneity within a continuum insulating 
medium representing an amorphous polymer phase.  
To this end, we introduce an off-lattice percolation algorithm which generates
arrangements of non-overlapping spheres in contact, starting from a random arrangement of particles \cite{SupMat}. 
We start by populating a cubic box of edge $L$ with $N$ spheres having centers drawn randomly from a uniform distribution. 
To each pair of overlapping spheres, we associate a compressed linear spring of unitary stiffness that connects the two particle centers and
minimize the following potential:
\begin{equation}
\label{V1}
\mathcal{V}=\sum_{i,j}\frac{1}{2}\Delta^2_{ij},
\end{equation}
where the sum runs over all pairs of overlapping spheres and
\begin{equation}
\label{V2} \Delta_{ij}=\left\{
\begin{array}{lll}
r_{ij}-D & \textrm{if} & r_{ij}<D \\
0         & \textrm{if} & r_{ij}\geq D,
\end{array}\right.
\end{equation}
where $r_{ij}$ is the distance between two sphere centers. To minimize $\mathcal{V}$, we use a gradient descent iterative algorithm 
applied to the subset of overlapping spheres. During the iteration, initially non-overlapping spheres remain untouched unless
they interfere with neighboring particles, in which case they are included in the summation of Eq.~\eqref{V1}. The iteration stops when
the pair that overlaps the most is such that $\vert\Delta_{ij}\vert/D\leq \delta$, where $\delta$ is a tolerance that we have fixed at $0.1$\%.
The final configuration is constituted by a dispersion of $N$ spheres within a cubic box of linear size $L$ that are either non-overlapping or in contact (with a $0.1$\% tolerance), so that the volume fraction of such spherical fillers is $\phi=\pi (N/L^3) D^3/6$.  

The configurations obtained by running the contact algorithm are characterized by fairly homogeneous dispersions of the conductive phase and 
represent CPCs in which micrometric fillers are dispersed into an amorphous polymer matrix.

\subsection{Aggregation algorithm}
\label{aggregation}
To simulate the effect of aggregation of the conducting phase, we adopt an algorithm applied to the configuration obtained by the
contact algorithm described above. The aggregation algorithm starts by identifying the component, denoted $\mathcal{C}$, formed by the 
largest number of particles in contact (for more information, see \cite{SupMat}). A particle $i$, randomly selected from the subset of particles not belonging to $\mathcal{C}$, is 
then subject to the following criterion. 
If its center is at a distance $r_{ij}$ smaller than $\lambda D$, with $\lambda\geq 1$, from the center of the closest particle, $j$, of the cluster, then $i$ is moved into 
contact with $j$ and the cluster $\mathcal{C}$ is updated. If instead $r_{ij}>\lambda D$, the particle $i$ can grow its own cluster, separated 
from $\mathcal{C}$, by following the same procedure. Once the aggregation algorithm is iterated and completed, some particles might overlap. 
We then rerun the contact algorithm to ensure that the final particle arrangement contains no overlapping spheres. 

The resulting dispersions of conductive fillers are characterized by regions of linear size $\sim \lambda D$ that are devoid of conductive 
spheres, which are instead segregated into regions of higher local concentration than the average $\phi$. For such types of randomly 
segregated distributions, the percolation threshold $\phi_c$ decreases as $\lambda$ increases.

\subsection{Volume expansion algorithm}
\label{volume}
The essential ingredient governing the phenomenology of the pyroresistive responses in CPCs is the uneven volume expansion of the
polymer and conducting phases during the heating of the composite. To focus on this aspect, we will not explicitly consider 
the thermal response of the polymer matrix, i.e., how the polymer expands as a function of temperature, but rather 
impose an isotropic strain $\varepsilon_0$ to the continuum matrix to simulate the effect of a polymer volume expansion,
whereas the spherical particles that constitute the conductive phase will be considered perfectly stiff.

We calculate the strain field within the composite and the resulting displacement of the spherical particles by using a meshless method
based on the reproducing kernel particle method of Ref.~\cite{Liu1995}. For more details about the algorithm, see \cite{SupMat}. One of the advantages of meshless shape functions over 
finite element ones is that the approximation is constructed entirely over nodes 
without the need for a tessellation or a mesh. This feature is crucial for domains with spheres in contact, where meshing between spheres 
is almost impossible. The price is a slight increase in computational costs because a meshless method requires a neighbor search between 
nodes and evaluation points in the continuum. Also, it requires a matrix inversion at each evaluation point. 
However, an appropriate space-partitioning data structure (k-d tree) minimizes the computational burden of neighbor search, and 
an iterative algorithm based on the Sherman-Morrison formula speeds up the matrix inversions \cite{Barbieri2012}.

\subsection{Resistor network}
\label{resistivity}
For each configuration of the conductive particles generated by the contact and aggregation algorithms described above, we construct 
a resistor network where the nodes
represent the centers of the spherical particles and assign to each pair of nodes of particles at contact a conductance of unit value.
We apply a unit voltage drop between two electrodes placed at the two opposite sides of the cubic simulation box and check if
a cluster of electrically connected nodes has terminal nodes in contact with the two electrodes. We then numerically solve the Kirchhoff equation
by matrix inversion to obtain the node voltages $V_i$. Since the edges linking two connected nodes have unitary resistance, the bond currents 
are given simply by $I_{ij}=V_i-V_j$.

The network equivalent current $I_\textrm{eq}$ is obtained by
summing over the currents of all bonds connected with the terminal nodes of one side of the simulation box. Because of Kirchhoff's current law, this sum
is equal to the sum of all bond currents of the nodes on the opposite side, so that the network equivalent resistance is $R_\textrm{eq}=1/I_\textrm{eq}$ 
and finally the resistivity is calculated from $\rho=R_\textrm{eq} L$ \cite{SupMat}.

\begin{figure*}[t]
\includegraphics[width=17 cm]{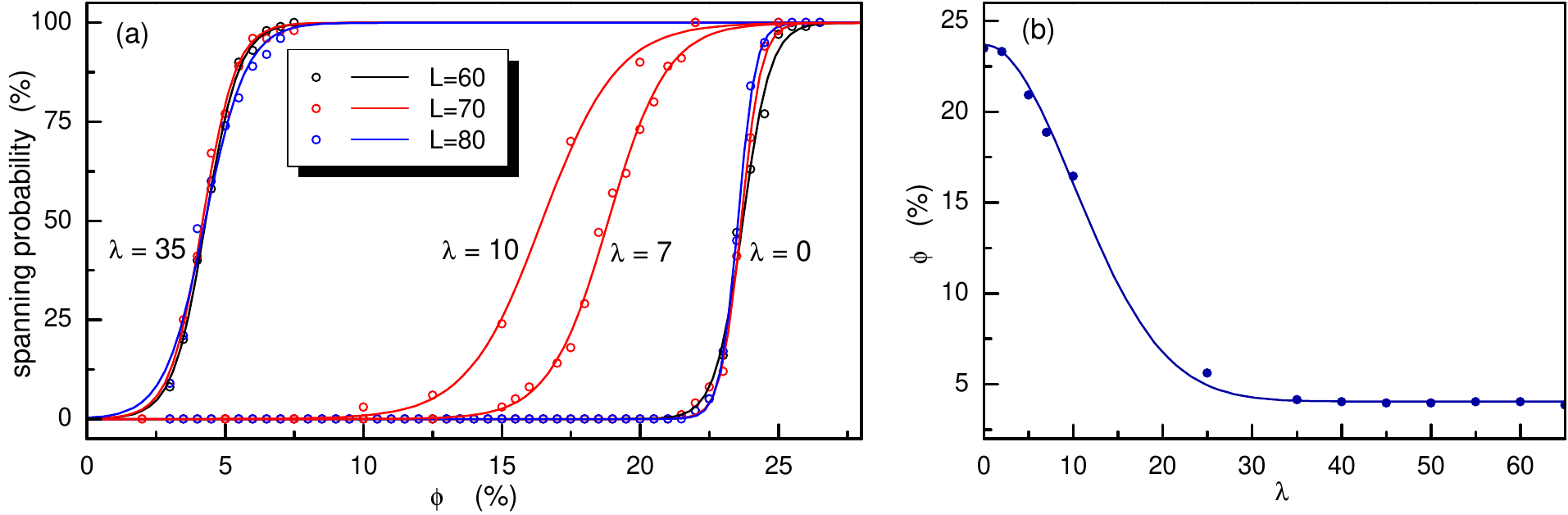}\\
\caption{Percolation threshold of homogeneous ($\lambda=0$) and segregated ($\lambda>0$) composites. (a) Spanning probability as 
a function of the filler volume fraction $\phi$ for different values of the aggregation length $\lambda$ in units of the sphere radius of the fillers. 
The critical volume fraction $\phi_c$ is determined by the spanning probability equal to $50$ \%. (b) The critical volume fraction $\phi_c$ as a 
function of $\lambda$. As the degree of segregation increases, the critical volume fraction decreases monotonously from $\phi_c\simeq 23.5$\% 
to $\phi_c\simeq 4$\%.}
\label{fig1}
\end{figure*}

\section{Results}
\label{results}

\begin{figure*}[t]
\includegraphics[width=17.5 cm]{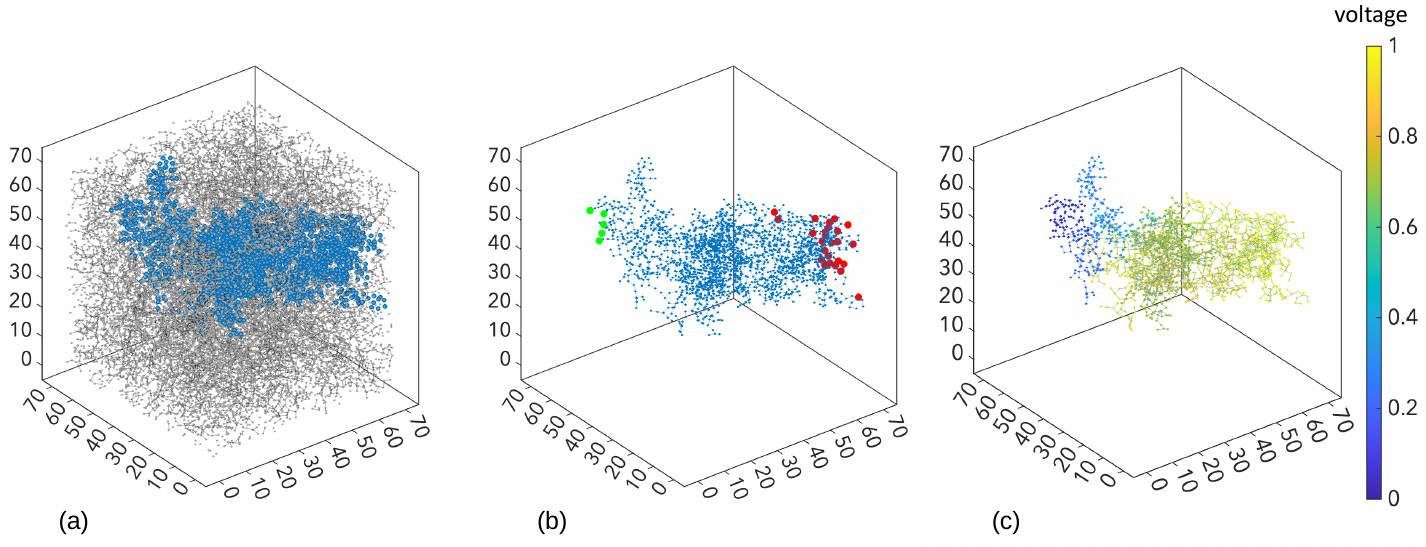}\\
\caption{Percolating component: (a) Network of contact particles. Dots are the vertices (sphere centers) and lines are the edges; 
the spheres forming the percolating cluster are highlighted in blue. (b) The percolating cluster with terminal nodes, where 
the red circles represent the ground nodes (voltage set to zero) and the green circles are terminal nodes with the voltage set to $1$. 
(c) Voltage distribution at the network nodes.}
\label{fig2}
\end{figure*}

\subsection{Percolation threshold}
\label{phic}
The electrical connectivity of our model composite is established only if there exists a system-spanning cluster formed by particles at contact.
To calculate the percolation threshold $\phi_c$ above which such a cluster exists, we run several configurations of the system for a fixed volume 
fraction of the conductive fillers and enumerate the number of times that a cluster of particles at contact spans the simulation box from one 
side to the opposite one. Figure~\ref{fig1}(a) shows the resulting spanning probability as a function of $\phi$ for fixed aggregation radii $\lambda$ 
obtained from $100$ configurations in a simulation box of linear size $L=70$ in units of the particle radius (red circles). To calculate 
the percolation threshold, we adopt the criterion that $\phi_c$ is the value of $\phi$ such that the spanning probability equals $50$ \% \cite{Rintoul1997}.
This can be obtained by fitting the data with a sigmoid function (solid lines) of the form:
\begin{equation}
\label{logistic} \frac{1}{2}\left( 1+\tanh\left[\frac{\phi-\phi_c(L)}{\Delta}\right] \right),
\end{equation}
where $\phi_c(L)$ is the percolation threshold for the simulation box of size $L$ and $\Delta$ is the width of the percolation transition.
In the case of homogeneous dispersions of particles, $\lambda=0$, we obtain $\phi_c\simeq 23.5$\%, which is comparable
to the critical volume fraction ($\simeq 20$ \%) of a random close packing of insulating and conductive spheres \cite{Ziff2017}. 
Increasing $\lambda$ systematically shifts the spanning probability towards smaller values of $\phi$, thereby leading to smaller percolation 
thresholds than the homogeneous case. This feature is systematically observed in numerical studies of segregated systems \cite{Kusy1977,He2004,Johner2009,Nigro2011} or other highly non-homogeneous dispersions of particles \cite{Nigro2013}.

Figure \ref{fig1}(b) shows the computed values of $\phi_c$ as a function of the aggregation radius $\lambda$.
Clearly, larger values of $\lambda$ entail smaller values of $\phi_c$, as is expected from the effect of segregation of the conductive phase.
In particular, we find that $\phi _c$ initially decreases with $\lambda$, eventually reaching the asymptotic value of $\phi_c\simeq 4$\% for $\lambda$
larger than about $40$. The reduction of the percolation threshold of over $80$\% 
indicates that the segregated percolating network is much more robust than homogeneous ones. 
Such robustness derives from the fact that, for a given volume fraction, the average 
number of contacts per particle increases as the conducting phase is increasingly segregated in narrower regions of the volume 
space \cite{Johner2009}.

In the following we will narrow our analysis to two specific scenarios: the homogeneous dispersion of particles ($\lambda=0$) and the highly 
segregated case achieved with $\lambda=35$. To assess the impact of the finite size of the simulation box on their percolation thresholds,
Fig. \ref{fig1}(a) shows the spanning probability calculated for $L=70$ alongside those computed for $L=60$ and $L=80$. These box sizes are large 
enough to prevent significant variations in $\phi_c(L)$. We find an overall change in the percolation threshold of only $0.5$\% for $\lambda=0$ and $4$\% for $\lambda=35$. 

\begin{figure}[t]
\includegraphics[width=8.5 cm]{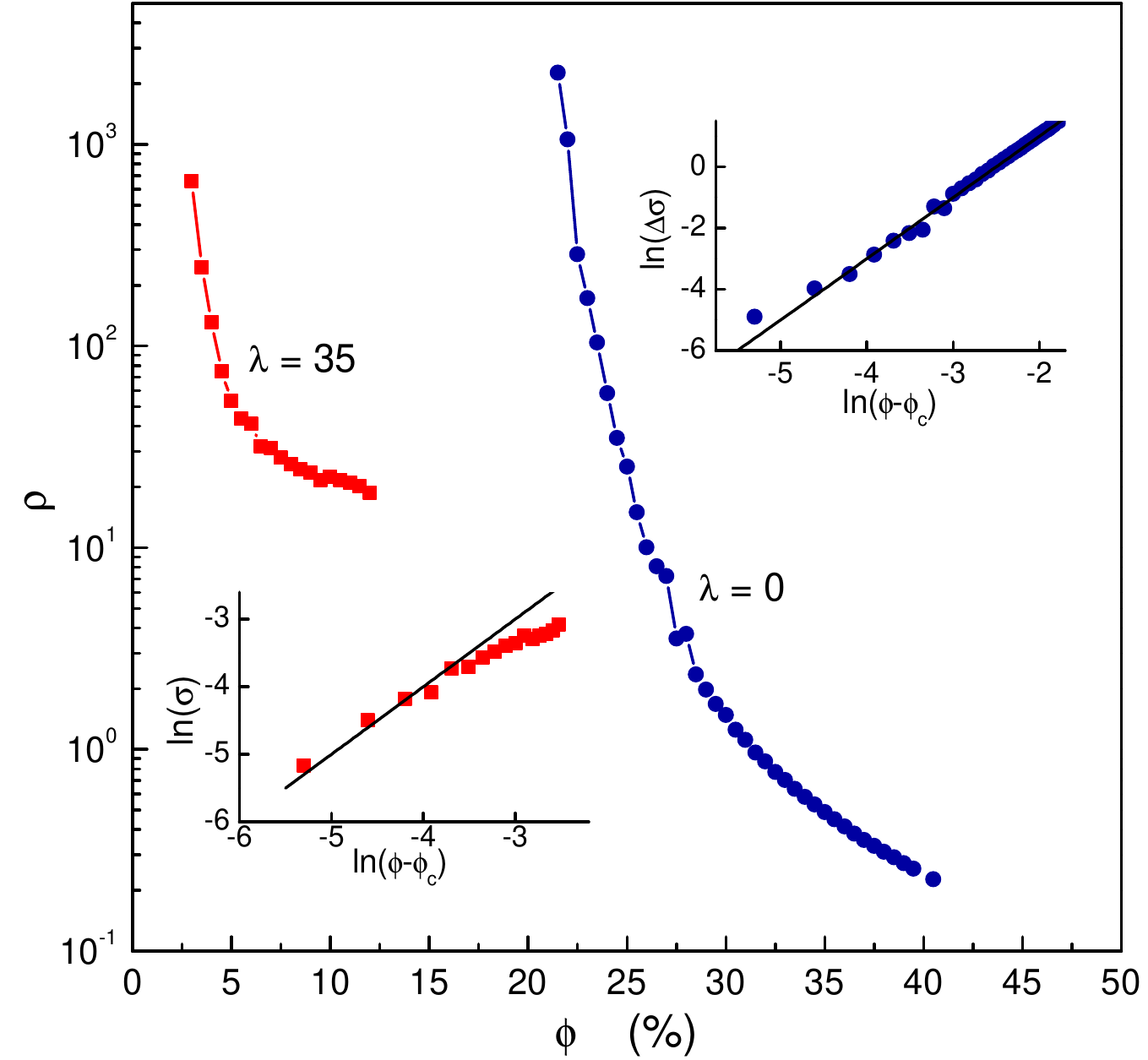}\\
\caption{Calculated resistivity $\rho$ obtained from numerical solutions of the Kirchhoff equation for homogeneous ($\lambda=0$) and 
highly segregated distributions of conductive fillers ($\lambda=35$). The insets highlight the power law behavior of the electrical conductivity 
$\sigma$. }
\label{fig3}
\end{figure}

\subsection{Resistivity}
\label{rho}

Figure~\ref{fig2}, shows the steps to calculate the network resistivity resulting from one realization of the spherical filler dispersion
obtained from the contact algorithm. The percolating connected component is represented by the set of blue dots in Fig.~\ref{fig2}(a), and the terminal nodes are represented in Fig.~\ref{fig2}(b) by the red and green circles. Figure~\ref{fig2}(c) shows the resulting node voltages $V_i$ obtained by the 
matrix inversion. 

Figure \ref{fig3} shows the composite resistivity calculated as a function of the spherical filler volume fraction $\phi$ for homogeneous 
dispersions ($\lambda=0$, filled circles) and a representative case of segregated composites obtained by the aggregation algorithm with aggregation distance equal to $\lambda=35$ (filled squares).
For each value of $\phi$, the box size is fixed at $L=70$ in units of the sphere radius and the resistivity data are the average over $100$ 
independent configurations. 
As seen from Fig.~\ref{fig3}, the resistivity of both types of composites increases monotonically as $\phi$ decreases, in qualitative agreement with
the percolative behavior $\rho\propto (\phi-\phi_c)^{-t}$, although the finite size of the simulation box entails finite
values of $\rho$ even for volume fractions smaller than the percolation thresholds of Fig.~\ref{fig1}. To mitigate the finite size effects and to get
a more quantitative analysis of the power-law behavior of the electrical transport, we plot in the insets of Fig.~\ref{fig3} the difference between the
conductivity computed at $\phi$ and that computed at the percolation threshold values obtained from Fig.~\ref{fig1}, $\Delta\sigma=\sigma(\phi)-\sigma(\phi_c)$ \cite{Johner2008}. From fitting to $\Delta\sigma\propto (\phi-\phi_c)^t$, the critical exponents for the homogeneous ($\lambda=0$) and segregated ($\lambda=35$) 
systems were found to be equal to $t=2.07\pm 0.03$ (in agreement with the universal value $t\simeq 2$) and $t\simeq 1$, respectively.

\subsection{Polymer expansion effects on the resistivity}
\label{expansion}
Having established how the type of dispersion of the conducting phase into the polymer matrix affects the resistivity behavior, we now turn to
addressing the effect of the polymer expansion on the transport properties of our model of CPCs. We are particularly interested in the
local expansion of the polymer matrix since this directly affects the connectivity of the conducting spherical particles. To understand why 
the local rather than the average polymer expansion matters, it suffices to realize that, ideally, any two particles initially at contact would be instantaneously 
separated by a perfectly homogeneous expansion of the polymer matrix. In this case, even an infinitesimal increase in the particle separations
would disrupt any connected component, thereby leading to an electrically insulating composite. In terms of a temperature-driven polymer expansion,
this situation would correspond to a conductor-insulator transition for any infinitesimal increase of $T$. In a more realistic modeling of CPCs, however,
the mismatch between the volume expansion of the conducting and insulating phases would build a highly heterogeneous stress field, which results
in a spatially varying strain field within the composite. In response to such a heterogeneous strain field, some particles initially at contact would
be separated (positive strain), whereas others would still be in contact (zero strain), or new connections might be realized from particles that
were initially slightly separated (negative strain). 

\begin{figure}[t]
\includegraphics[width=8.5 cm]{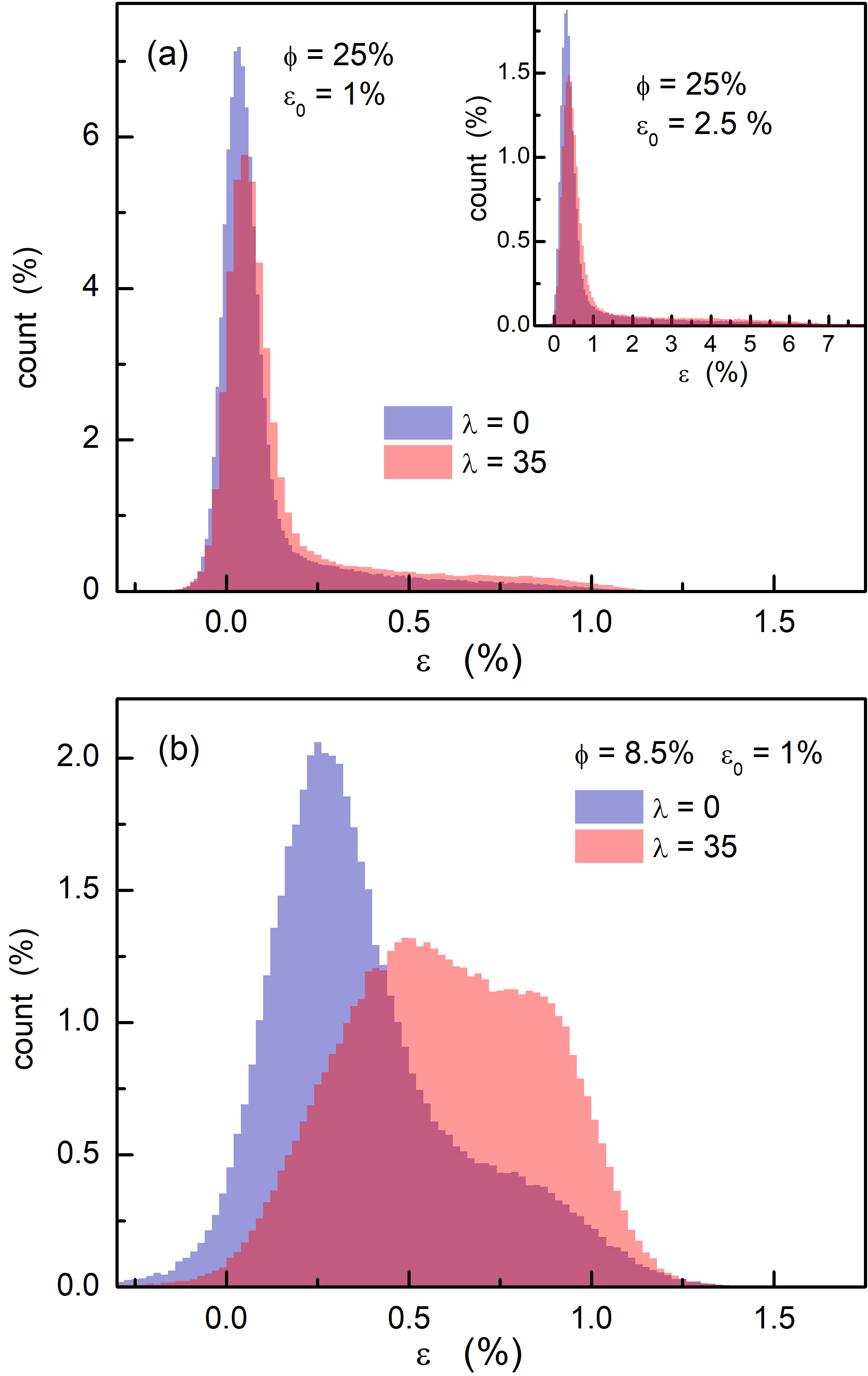}\\
\caption{Distribution function of the maximum principal strain $\varepsilon$ obtained from the numerical solution of the thermoelastic 
equations for homogeneous
($\lambda=0$) and highly segregated ($\lambda=35$) dispersions of the conductive spherical fillers in the insulating polymer matrix. 
The thermal strain is set equal to 
$\varepsilon_0=1$\%. (a) The volume fraction is equal to $\phi=25$\%, for which both cases are above their respective percolation threshold. (b)
$\phi=8.5$\%. For this value of $\phi$, only the homogeneous system is below the percolation threshold.}
\label{fig4}
\end{figure}

This situation is illustrated in Fig.~\ref{fig4}, which shows the distribution of the strain field calculated by imposing a strain of pure polymer
equal to $\varepsilon_0=1$\%. For both homogeneous ($\lambda=0$) and segregated ($\lambda=35$) dispersions of fillers with $\phi=25$\%
(Fig.~\ref{fig4}(a)),
the strain distribution is strongly peaked at around $\varepsilon=0.03-0.05$\% with a small shoulder extending up to $\varepsilon\simeq 1$\% and
a significant contribution (about $16$\% and $10$\% of the total weight) from negative strains down to about $-0.1$\%. Although the two
distributions are similar for $\phi=25$\%, the strain distribution of the segregated system exhibits a smaller peak compared 
to the homogeneous system and slightly greater weight at larger $\varepsilon$. This trend also persists at larger values of the 
thermal strain, as shown in the inset of Fig.~\ref{fig4}(a) for $\varepsilon_0=2.5$\%. Notably, in this case the strain
distribution extends significantly beyond $2.5$\%, with the segregated systems displaying a larger weight in this range. 
This translates into a higher average local strain for the segregated system ($\bar{\varepsilon}=1.5$\%) compared to the 
homogeneous system ($\bar{\varepsilon}=1.1$\%).

At a volume fraction of only $8.5$\% [Fig.~\ref{fig4}(b)], the strain distribution of both types of systems gets significantly broadened and skewed toward 
positive values of $\varepsilon$ because the polymer
expansion is less restrained by such a small concentration of rigid fillers (although a small contribution from negative strains is still present). 
However, the $\varepsilon$-distribution of the segregated system displays a broad plateau extending from $\varepsilon\sim 0.5$\% to
$\varepsilon\sim 1$\%, whereas the distribution for the homogeneous system is significantly smaller in the same range. The overall behavior
shown in Fig.~\ref{fig4} suggests, therefore, that, although the segregated system is more robust against a change of filler content compared 
to the homogeneous one, as seen from the smaller value of $\phi_c$, the response to the thermal strain appears to be comparable 
for the two types of filler dispersion, with actually the segregated system being more responsive to the thermal strain than the homogeneous one.

\begin{figure}[t]
\includegraphics[width=8.5 cm]{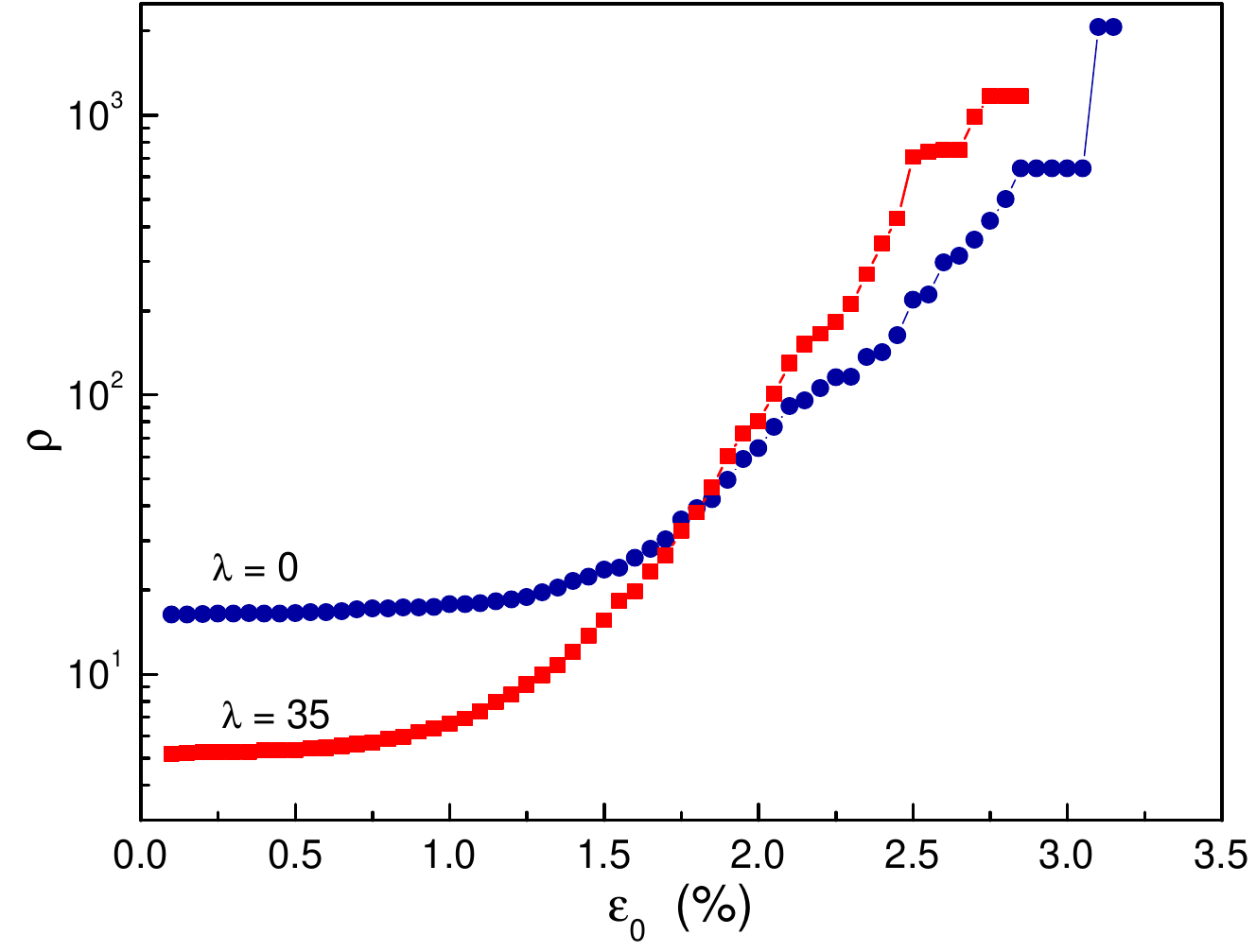}\\
\caption{Resistivity $\rho$ calculated as a function of the thermal strain for homogeneous ($\lambda=0$) and highly segregated ($\lambda=35$)
dispersions of the conductive fillers with $\phi=25$\%. Note that for the segregated system $\rho$ starts to already increase for 
$\varepsilon_0\gtrsim 1$\%, despite being far above the percolation threshold.}
\label{fig5}
\end{figure}

The electrical transport response to the thermal strain is
shown in Fig.~\ref{fig5}, where we plot the computed resistivity $\rho$ as a function of the thermal
strain $\varepsilon_0$ for both the homogeneous and segregated dispersions of fillers with $\phi=25$\%. 
For values of the thermal strain up to about $1$\%, $\rho$ undergoes only a moderate change for both
$\lambda=0$ and $\lambda=35$, with the segregated composite showing slightly enhanced resistivity compared 
to the homogeneous one.  This modest change in resistivity for $\varepsilon_0=1$\% is coherent with the strain field 
distribution depicted in Fig.~\ref{fig4}(a).
Upon increasing $\varepsilon_0$ beyond $1$\%, the segregated system exhibits a significant rise in the resistivity, 
eventually becoming an insulator for $\varepsilon_0> 2.85$ \%,
whereas the resistivity of the homogeneous dispersions starts to increase only for 
$\varepsilon>1.5$\% with the transition to the insulating state at $\varepsilon_0\simeq 3.15$\%. 
Furthermore, the resistivity increase of the segregated system is steeper than that of the homogeneous one.
This is even more relevant if we consider that at $\phi=25$\% the filler loading of the segregated system is far above the percolation
threshold at $\phi_c=4$\%, implying, therefore, that the percolating network is more fragile under a polymer volume expansion
when the fillers are segregated into the polymer. 

\begin{figure}[t]
\includegraphics[width=8.5 cm]{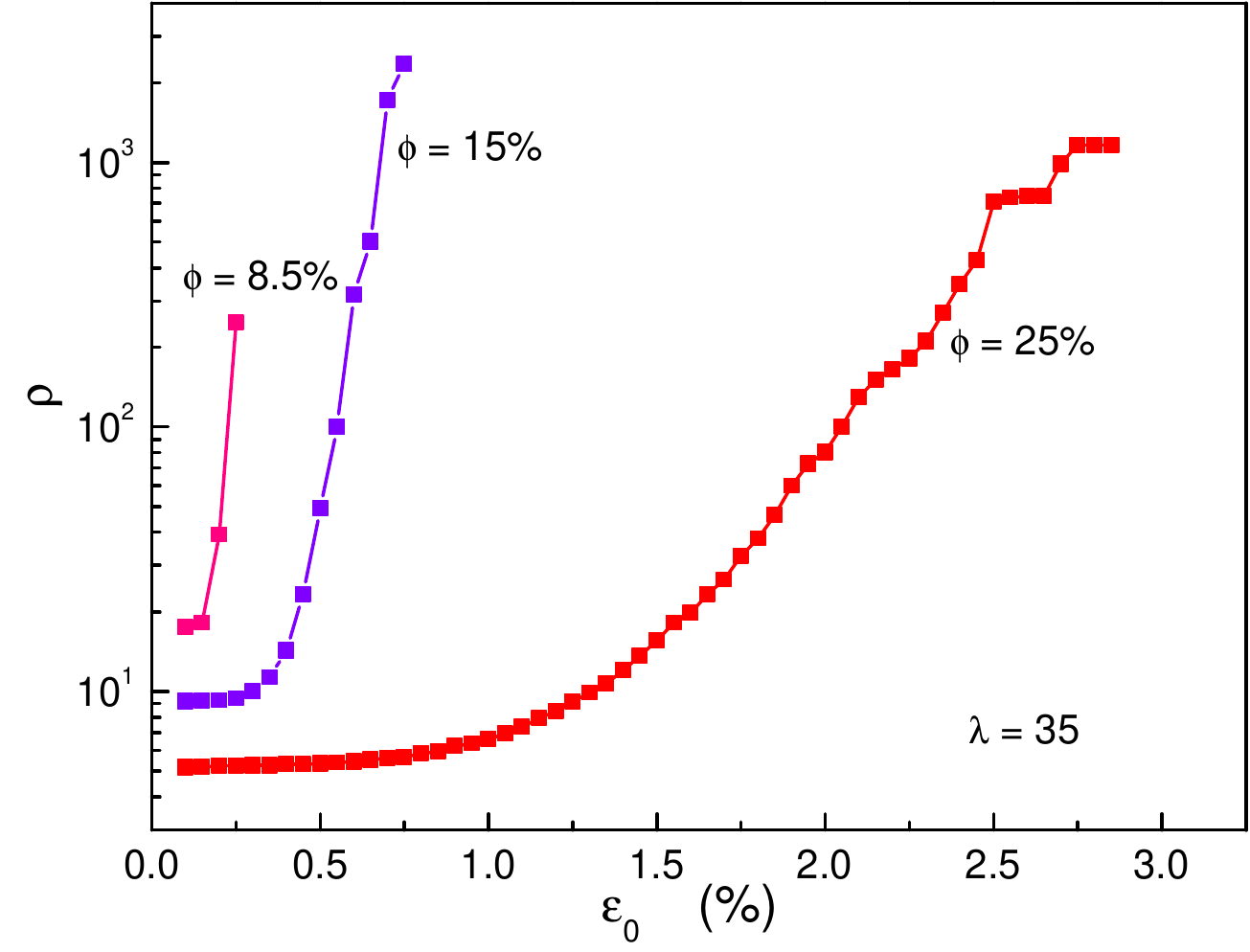}\\
\caption{Resistivity $\rho$ calculated as a function of the thermal strain a highly segregated ($\lambda=35$)
dispersion of the conductive fillers for different values of the nominal volume fraction $\phi$. As the filler content approaches the percolation
threshold at $\phi_c\simeq 4$\%, the thermal strain at which the resistivity increases shifts to smaller values.}
\label{fig6}
\end{figure}

The strain distribution function shown in Fig.~\ref{fig4}(b) suggests that the conductive
network can be disrupted for smaller thermal strains when the filler content approaches the percolation threshold. This is confirmed in Fig.~\ref{fig6}, where we show the calculated resistivity as a function
of the thermal strain for the segregated composite with $\phi= 25$\%, $15$\%, and $8.5$\%. The switching value of the strain becomes smaller as
$\phi$ approaches $\phi_c=4$\%, demonstrating that the resistivity response can be tuned by the filler loading. 

This phenomenon has been observed in CPCs incorporating micrometric metallic fillers \cite{Kar2010,Kar2012,Xi2004,Mitsuhiro2005,Rybak2010,Kono2012,Zhang2014,Xu2022},
wherein a decrease in metallic content leads to lower values of the switching temperature $T_0$. For instance, in CPCs containing
Ag and Cu particles, the volume expansion measured at $T_0$ implies a thermal strain in the range 
$0.4$\%-$0.6$\% \cite{Kar2010,Rybak2010}. This is  below the large strains accompanying the polymer structural 
transition and is comparable to those of Figs.~\ref{fig5} and \ref{fig6}.

\begin{figure}[t]
\includegraphics[width=8.5 cm]{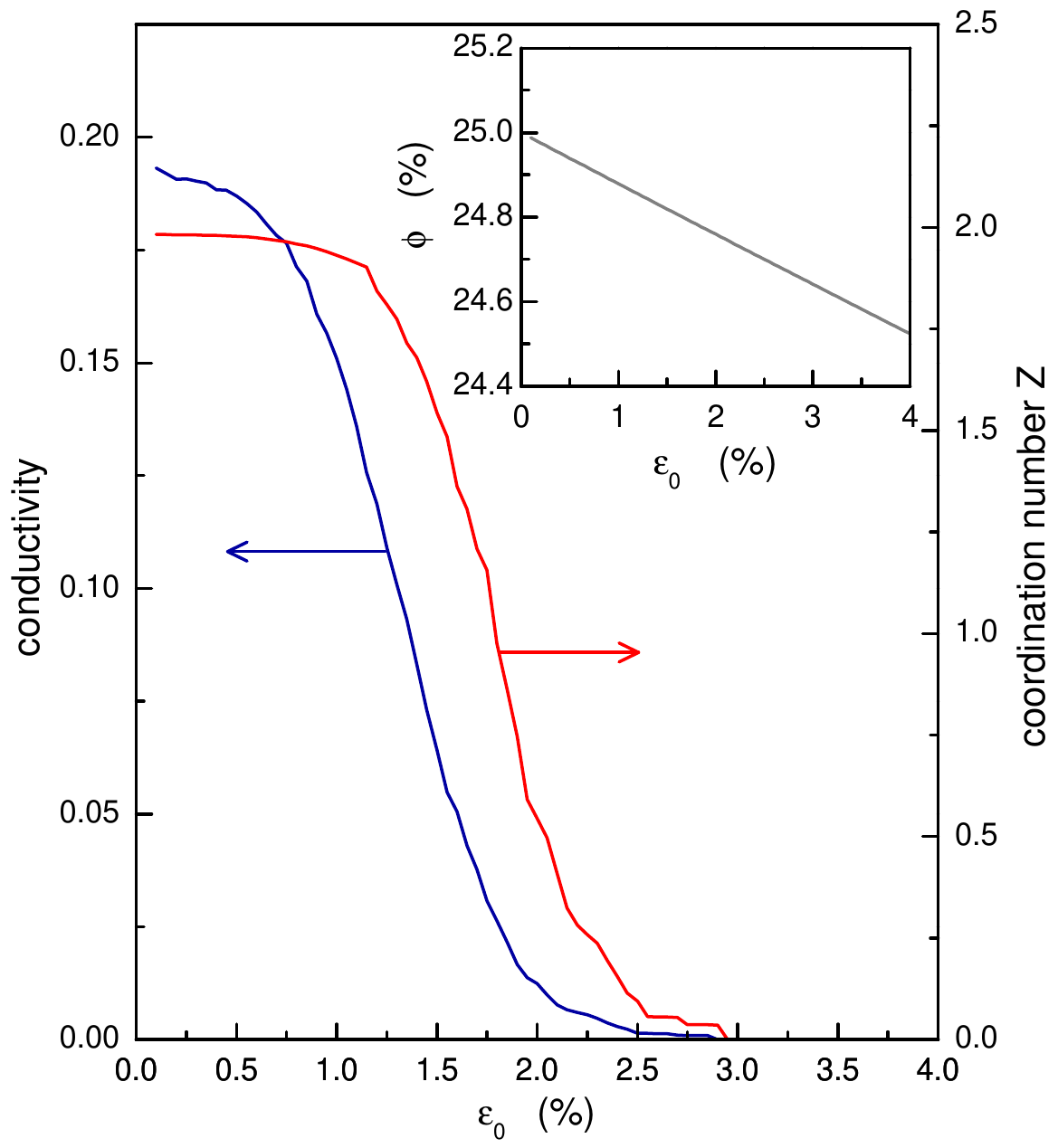}\\
\caption{Average coordination number $Z$ and the electrical conductivity $\sigma$ calculated for an highly segregated system 
($\lambda=35$) as a function of the thermal strain $\varepsilon_0$. The nominal volume fraction is equal to $25$\%. 
Inset: the corresponding effective volume fraction plotted against $\varepsilon_0$.}
\label{fig7}
\end{figure}

To better understand the interplay between the local strain and the network connectivity, we plot in Fig.~\ref{fig7} the average coordination number
$Z$ as a function of the thermal strain $\varepsilon_0$ for the segregated system at $\phi=25$\%. The quantity $Z$ gives the average number of particles 
at contact with any given particle. It provides a less coarse-grained measure of the local connectedness 
than $\phi$, and for this reason, it represents a better tool to analyze how local changes of the percolating network influence the conductivity. 
This is particularly evident in Fig.~\ref{fig7}, which shows that $Z$ rapidly decreases from $Z\simeq 2$ at $\varepsilon_0=0$\% to $Z=0$ for $\varepsilon_0\simeq 3$\%, thus paralleling the decrease of the conductivity with $\varepsilon_0$, whereas $\phi$ diminishes of only $\sim 1.6$\%
in the same range (inset of Fig.~\ref{fig7}). 

The concomitant vanishing of $Z$ and $\sigma$ in Fig.~\ref{fig7} differs drastically from what is expected when a conductor-insulator system is driven 
towards the percolation threshold $\phi_c$ by lowering the amount of the conducting fillers. In that case, at the percolation threshold, the conductivity
vanishes, whereas the average coordination number attains a finite (critical) value $Z_c$, which depends on the local connectivity properties. 
For example, $Z_c\simeq 2.8$ for random distributions of penetrable
spheres or $Z_c\simeq 1$ for penetrable particles of high aspect-ratios such as slender spherocylinders.
In Fig.~\ref{fig7} we see instead that the transition to the insulating state is accompanied by the average coordination number
going to zero, suggesting that the effect of the thermal expansion is that of a global disruption of the percolating network
(basically, all links get disconnected at the same time) as opposed to the gradual disruption when the filler volume fraction is reduced, where
the weakest pathways of the network are disconnected first.

\section{Discussion and conclusions}
\label{concl}

In this paper, we have shed some light on the microscopic mechanism of the pyroresistive response of CPCs by numerically analyzing
the coupling between the thermoelastic response of the composite and the electrical transport properties. We have shown that for CPCs
that can be modeled as percolating networks formed by contacts between conductive fillers, the local strain built up by an isotropic thermal expansion
of the pure polymer governs the electrical connectedness of the conductive phase. The difference between the thermal volume expansion
of the insulating medium (the polymer) and the conductive fillers induces a highly heterogeneous local strain field, which can separate
particles that were initially in contact, eventually leading to the breakdown of the conductive network and the consequent divergence of the resistivity.
This conductor-insulator transition depends on the volume fraction $\phi$ of the conductive phase, such that the closer $\phi$ is to the percolation threshold, the smaller the thermal strain at which the resistivity diverges.
However, in sharp contrast to the percolating behavior of transport when $\phi\rightarrow \phi_c$, we find in this case that the divergence of the 
resistivity is driven by the mean number of particles at contact going to zero, indicating a sudden disruption of the conductive network
rather than a gradual one.
Furthermore, we have shown that the type of dispersion of the conductive fillers can have important effects on the pyroresistive response.
In particular, highly heterogeneous distributions of the conductive phase, as typically found in segregated CPC systems, are more fragile
under a polymer volume expansion than homogeneous ones, even though segregated composites typically have much smaller percolation thresholds.

Although our paper focuses on a simple model of CPCs in which the polymeric phase is amorphous and the conductive fillers are identical spheres,
we are not aware of other analyses in which the effects of the local volume change of the polymer are explicitly coupled to the electrical connectivity of the
conductive particles. This has allowed us to compute the local strain fields, which are the ultimate drivers of the filler displacements within the composite,
and to quantify the change in resistivity of the percolating network. More realistic models of CPCs would, however, consider several 
characteristics of real materials whose effects on the pyroresistance we have chosen to postpone to later studies. For example, as mentioned in 
the Introduction, the size of the conductive fillers might play an important role in the phenomenology of the PTC effect, as it is well-known that the 
electrical connectivity between nanometric conductive particles is governed by the tunneling decay length. In this case, the connectivity can extend 
beyond first neighbors and open conductive pathways that can influence the pyroresistive response. Modelling the polymer 
as a homogeneous continuous medium, as done in this paper, does not conform with those CPCs in which the polymer is partially crystalline. 
Furthermore, the volume expansion of the crystalline phase upon heating differs from that of the amorphous phase, thus influencing 
the onset of the PTC effect.
Finally, another feature worth studying concerns the thermal conductance of the conductive phase. In fact, metallic fillers have higher thermal conductance than carbon-based ones and can, therefore, induce a stronger local volume expansion of the polymer.

For a more theoretical understanding of the nature of the strain-induced conductor-insulator transition, it would be interesting 
to study in more detail the evolution of the conductivity as a function of the number of particles at contact when the system 
is under a thermal strain. To this end, a finite size analysis could shed light on whether the transition is sharp but still of
second-order or, instead, the system undergoes a discontinuous (i.e., first-order) transition.

In conclusion, the present paper represents an attempt to understand the microscopic mechanism 
at the origin of the pyroresistive behavior of CPCs. Our results may guide the designing of optimal CPC pyroresistors and stimulate further numerical simulations to improve our understanding of the PCT effect.

\renewcommand{\thefigure}{S\arabic{figure}}
\renewcommand{\theequation}{S\arabic{equation}}
\renewcommand{\thetable}{S\arabic{table}}
\renewcommand{\thesection}{S\arabic{section}}
\setcounter{section}{0}
\setcounter{equation}{0}
\setcounter{figure}{0}
\setcounter{table}{0}

\newcommand{\erf}{\operatorname{erf}}
\newcommand{\atand}{\operatorname{atan2}}
\newcommand{\x}{\mathbf{x}}
\newcommand{\X}{\mathbf{X}}
\newcommand{\U}{\mathbf{u}}
\newcommand{\UU}{\mathbf{U}}
\newcommand{\that}{\mathbf{\hat{t}}}
\newcommand{\nhat}{\mathbf{\hat{n}}}
\newcommand{\bhat}{\mathbf{\hat{b}}}
\newcommand{\phit}{\tilde{\mathbf{\phi}}}
\newcommand{\R}{\mathbf{R}}
\newcommand{\rr}{\mathbf{r}}
\newcommand{\Heav}{\mathcal{H}}
\newcommand{\de}{\mbox{d}}
\newcommand{\dep}[2]{\dfrac{\partial #1}{\partial #2}}
\newcommand{\ddep}[2]{\dfrac{\partial^2 #1}{\partial #2^2}}
\newcommand{\ddmp}[3]{\dfrac{\partial^2 #1}{\partial #2 \partial #3}}
\newcommand{\deT}[2]{\dfrac{\de #1}{\de #2}}
\newcommand{\ddeT}[2]{\dfrac{\de^2 #1}{\de #2^2}}
\newcommand{\sign}{\operatorname{sign}}
\newcommand{\dO}{\partial \Omega}
\newcommand{\fl}[1]{\left \lfloor #1 \right \rfloor}
\newcommand{\ce}[1]{\lceil #1 \rceil}
\newcommand{\saw}[1]{\{ #1 \}}
\newcommand{\sxinf}{\sigma_x^{\infty}}
\newcommand{\syinf}{\sigma_y^{\infty}}
\newcommand{\uL}{\mathbf{u}_L}
\newcommand{\T}[1]{\boldsymbol{#1}}
\newcommand{\Grad}{\nabla_0}
\newcommand{\Div}[1]{\nabla_0 \cdot #1}
\newcommand{\PKF}{\mathbf{P}}
\newcommand{\PKS}{\mathbf{S}}
\newcommand{\E}{\mathbf{E}}
\newcommand{\F}{\mathbf{F}}
\newcommand{\eye}{\mathbf{I}}
\newcommand{\D}{\mathbf{d}}
\newcommand{\h}[2]{ \dep{u_{#1}}{X_{#2}} }
\newcommand{\PSI}{\boldsymbol{\Psi}}
\newcommand{\KT}{\mathbf{K}_T}
\newcommand{\FintT}{\mathbf{F}^{(i) \ T}}
\newcommand{\Fint}{\mathbf{F}^{(i)}}
\newcommand{\FextT}{\mathbf{F}^{(e) \ T}}
\newcommand{\Fext}{\mathbf{F}^{(e)}}
\newcommand{\defmap}{\boldsymbol{\varphi}}
\newcommand{\Body}{\mathcal{B}}
\newcommand{\Real}{\mathbb{R}}
\newcommand{\Tess}{\mathcal{T}}
\newcommand{\ckso}{\Gamma_c^0}
\newcommand{\cksoi}{\Gamma_{c_{i}}^0}
\newcommand{\cks}{\Gamma_c}
\newcommand{\Int}{\mathcal{W}}
\newcommand{\Ext}{\mathcal{P}}
\newcommand{\Pot}{\mathit{\Pi}}
\newcommand{\sph}{\mathcal{S}}
\newcommand{\M}{\mathsf{M}}
\renewcommand{\P}{\mathbf{P}}
\newcommand{\disk}{\mathcal{D}_\rho}
\newcommand{\p}{\mathbf{p}}

\widetext{

\section*{Pyroresistive response of percolating conductive polymer composites: Supplemental Material}

\section{Contact Algorithm}
\label{sec:ContactAlgorithm}
The algorithm starts with $N$ spheres having centres drawn randomly from a uniform distribution. 
Let us call their coordinates $\rr_i\quad i=1,\ldots,N$ and the spheres' radii $R_i\quad i=1,\ldots,N$ (figure \ref{fig:Potential}).
This arrangement of particles likely results in overlapping spheres $\mathcal{O} = \left\{ (i,j) : \frac{r_{ij}}{R_i+R_j}<1 \right\}$, with $\rr_{ij} =  \mathbf{r}_i-\mathbf{r}_j$ and $r_{ij} = \|\rr_{ij}\|$ (figure \ref{fig:Potential}).  To compute $r_{ij}$ in a fast manner, we use a range search through a \emph{k-d tree} algorithm.

We treat each pair $(i,j) \in \mathcal{O}$ of overlapping spheres as connected by a compressed linear spring of unitary stiffness along the unit vector $\hat{\mathbf{r}}_{ij} = \rr_{ij}/r_{ij}$ (figure \ref{fig:ContactIterations}).  Such springs have no tensile stiffness (figure \ref{fig:Potential}), meaning that once particles are apart, they remain separated (figure \ref{fig:ContactIterations}).

With $\rr = \left[\begin{array}{cccc}
    \rr_1 & \rr_2 & \ldots & \rr_N 
\end{array}\right]^T$, we seek to minimise the following potential:
\begin{equation}
  \rr = \mathrm{argmin} \mathcal{V} \qquad \mathcal{V}= \sum_{(i,j)\in \mathcal{O}}\,\frac{1}{2}\, V_{ij} \qquad V_{ij} = \begin{cases}
    {g_{ij}}^2 & g_{ij}<0 \\
    0 & g_{ij} \geq 0
    \end{cases}
\end{equation}
with $g_{ij}$ being the \emph{gap function}
\begin{equation}
    g_{ij} = r_{ij}-\left(R_i+R_j\right)<0
\end{equation}

To minimize $\mathcal{V}$, we use a gradient descent iterative algorithm. 
This approach has the considerable advantage of being an explicit scheme, hence extremely fast. 
\begin{equation}
    \rr^{n+1} = \rr^{n} - \gamma\,\nabla\mathcal{V}
    \label{eq:gradient_descent}
\end{equation}
where the gradient $\nabla\mathcal{V} = \dep{\mathcal{V}}{\rr}$ is
\begin{equation}
    \nabla \mathcal{V} = \sum_{(i,j)\in \mathcal{O}}\, g_{ij}\, \hat{\rr}_{ij}
\end{equation}
We choose $\gamma=0.4$. From numerical experiments, this particular value of $\gamma$ proved to be an optimal compromise between stability and convergence rate. The algorithm stops when the pair that overlaps the most, has an overlap that is within a certain threshold:
\begin{equation}
   \min_{(i,j)\in \mathcal{O}}\, |{g_{ij}}^*|<\mathrm{tol} \qquad  {g_{ij}}^* = \frac{r_{ij}}{R_i+R_j} - 1
   \label{eq:criterion}
\end{equation}
where $\mathrm{tol}$ is a tolerance fixed at $0.1\%$.

The algorithm brings initially overlapping particles to be in contact with no overlap: non-overlapping particles are untouched unless, of course, they interfere with neighbouring particles during the iterations (figure \ref{fig:ContactIterations}).
\begin{figure}[htp]
    \centering
    \includegraphics[width=0.7\textwidth]{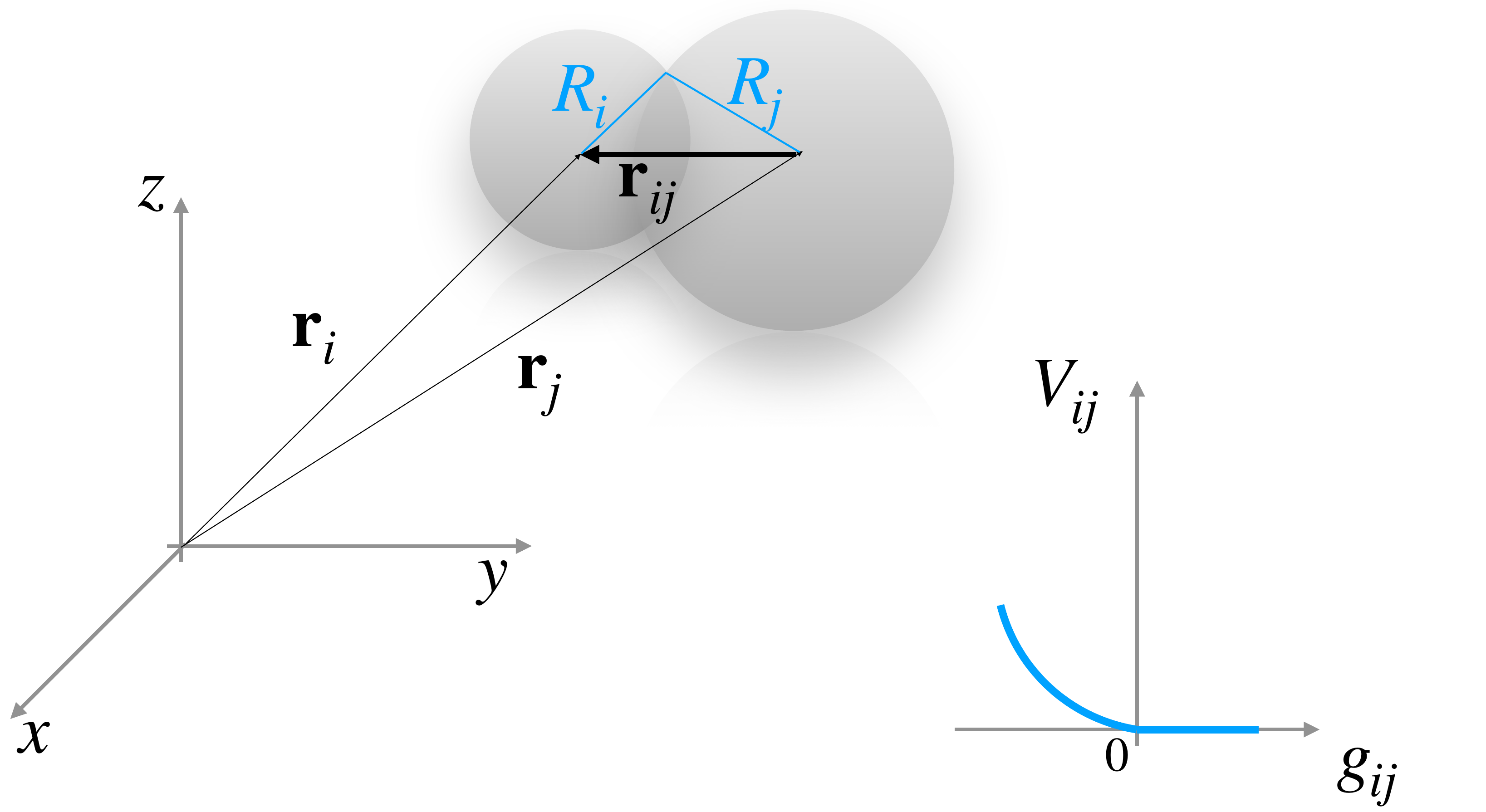}
    \caption{Positions $\mathbf{r}_i$ and $\mathbf{r}_j$ of two overlapping spheres of radii respectively $R_i$ and $R_j$. $V_{ij}$ is the potential of the compressive spring as a function of $g_{ij}=r_{ij}-(R_i+R_j)$.}
    \label{fig:Potential}
\end{figure}
\begin{figure}[htp]
    \centering
    \includegraphics[width=0.7\textwidth]{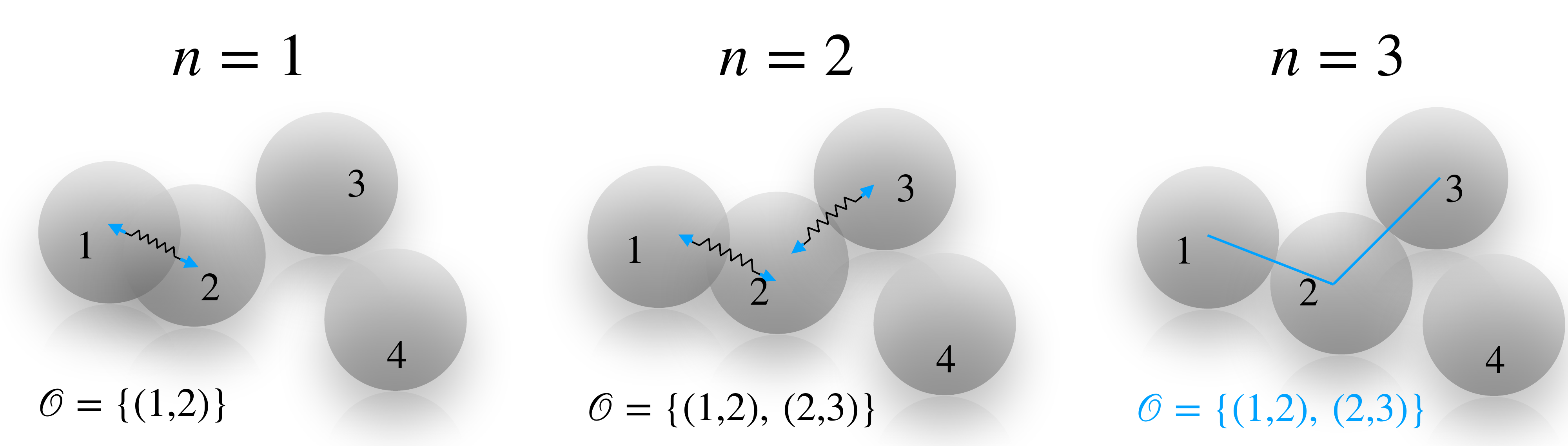}
    \caption{Contact algorithm: in this example, four spheres are initially present at iteration $n=1$, with spheres $1$ and $2$ overlapping. A spring displaces $1$ and $2$ reducing their overlap, but causing $2$ to overlap with $3$ (iteration $n=2$). Two elastic springs push apart the two pairs $(1,2)$ and $(2,3)$. Finally, at the last iteration $n=3$, pairs $1$ and $2$ and $2$ and $2$ are brought to only touching each other. Particle $4$ is left untouched, as it does not overlap with any spheres during the iterations.}
    \label{fig:ContactIterations}
\end{figure}

\section{The aggregation algorithm}
\label{sec:AggregationAlgorithm}
Firstly, we build a graph $\mathcal{G}$ from the final set $\mathcal{O}$ of contact pairs satisfying criterion \eqref{eq:criterion}. We then compute $\mathcal{G}$'s connected components, and select the one with the largest number of particles. Let us call $\mathcal{C}$ such cluster (figure \ref{fig:aggregation}) and $n_C$ the number of its particles with center coordinates $\mathbf{r}_1, \, \mathbf{r}_2, \, \ldots , \mathbf{r}_{n_{\mathcal{C}}}$.
Let $d_{\mathcal{C}}(\mathbf{r})$ be an Euclidean distance field from $\mathcal{C}$ (figure \ref{fig:aggregation}) such that
\begin{equation}
    d_{\mathcal{C}}(\mathbf{r}) = \underset{i=1,\ldots, n_{\mathcal{C}}}{\mathrm{min}} \  d_i(\mathbf{r}) \qquad d_i(\mathbf{r}) = |\mathbf{r}-\mathbf{r}_i|
\end{equation}

We then iterate over all the remaining particles in $\mathcal{G}$ not belonging to $\mathcal{C}$. We select a particle with center $\mathbf{r}_j$ if its distance $d_{\mathcal{C}}(\mathbf{r}_j)$ is less than $\lambda\,R$, with $R$ being the radius of the particle (figure \ref{fig:aggregation}). We assume a mono-disperse distribution of particles.

The particle is attracted to $\mathcal{C}$ according to the following optimization algorithm with a (non-linear) constraint that the particle does not overlap with any spheres in cluster $\mathcal{C}$:
\begin{equation}
   \mathbf{r}_j = \underset{\mathbf{r}}{\mathrm{argmin}}\ \mathbf{d}_{\mathcal{C}}(\mathbf{r}) \quad \mathrm{such \ that} \quad \mathbf{d}_{\mathcal{C}}(\mathbf{r}) \geq 2\,R \label{eq:optimization}
\end{equation}
If the optimization is successful, then the position $\mathbf{r}_j$ of the particle is updated and $\mathbf{d}_{\mathcal{C}}(\mathbf{r}_j)=2\,R$.
In any case, whether the initial $d_{\mathcal{C}}(\mathbf{r}_j) \leq \lambda\,R$ or otherwise, whether the optimization \eqref{eq:optimization} is successful, we grow $\mathcal{C}$ with particle of center $\mathbf{r}_j$. In this way, we allow isolated particles to grow their own separate cluster from the largest one (figure \ref{fig:aggregation2}).
Once the aggregation algorithm is completed, some particles might overlap. We then run the contact algorithm to $\mathcal{C}$ in section \ref{sec:ContactAlgorithm} again to make sure that the final arrangement of particles and contains no overlapping spheres.

\begin{figure}[htp]
    \centering
    \includegraphics[width=0.7\textwidth]{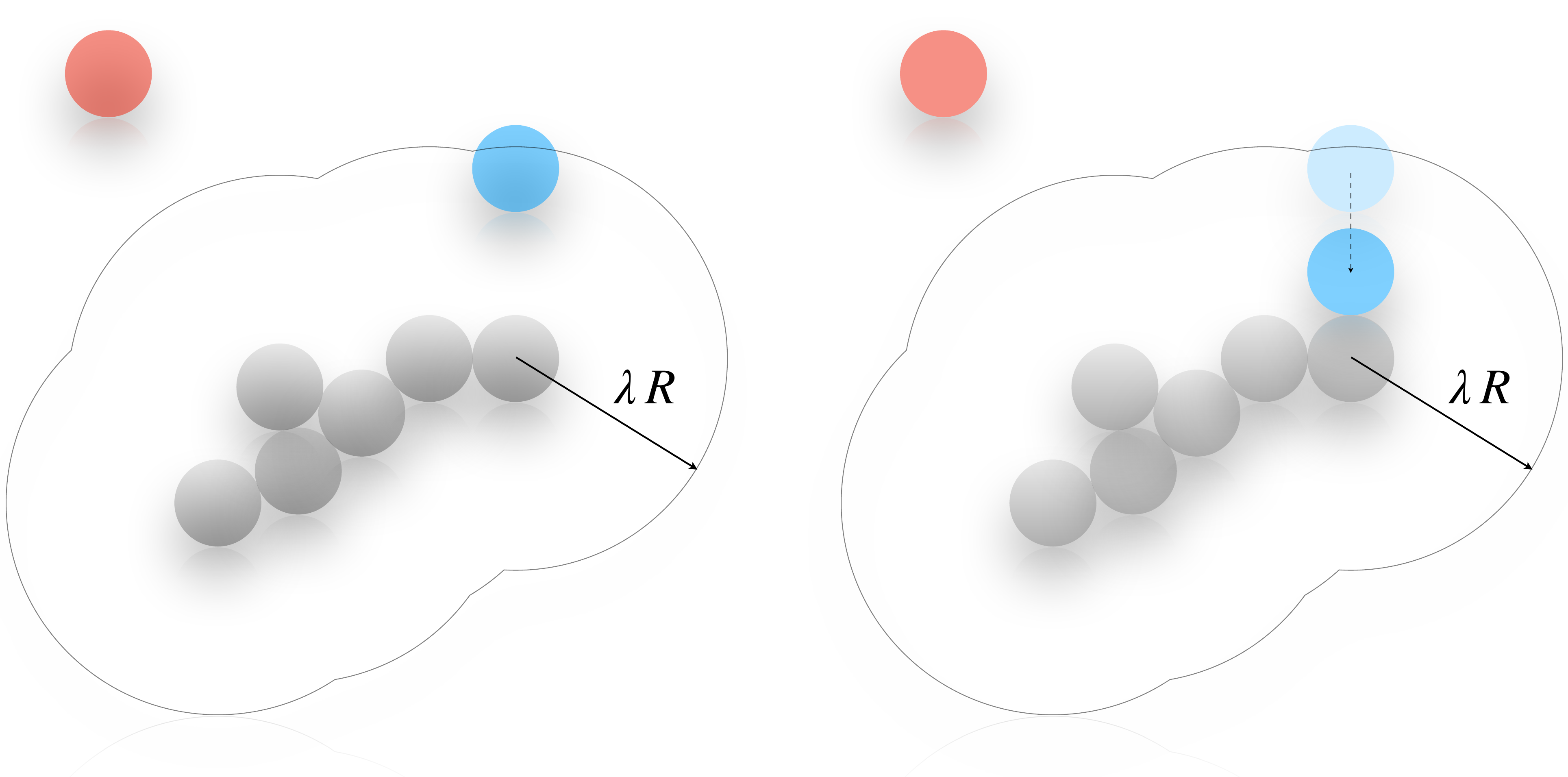}
    \caption{Aggregation algorithm: the first step is finding the largest connected component, in this case the cluster of grey particles; all the other particles are attracted if their center's distance $d_{\mathcal{C}}$ from this largest cluster is less than $\lambda\,R$ (black continuous line), with $R$ being the radius of the particle. In this example, the blue particle is attracted, while the red particle is not. The attraction to the cluster is carried out by minimizing the cluster-particle distance subjected to the no-overlap constraint.}
    \label{fig:aggregation}
\end{figure}
\begin{figure}[htp]
    \centering
    \includegraphics[width=0.7\textwidth]{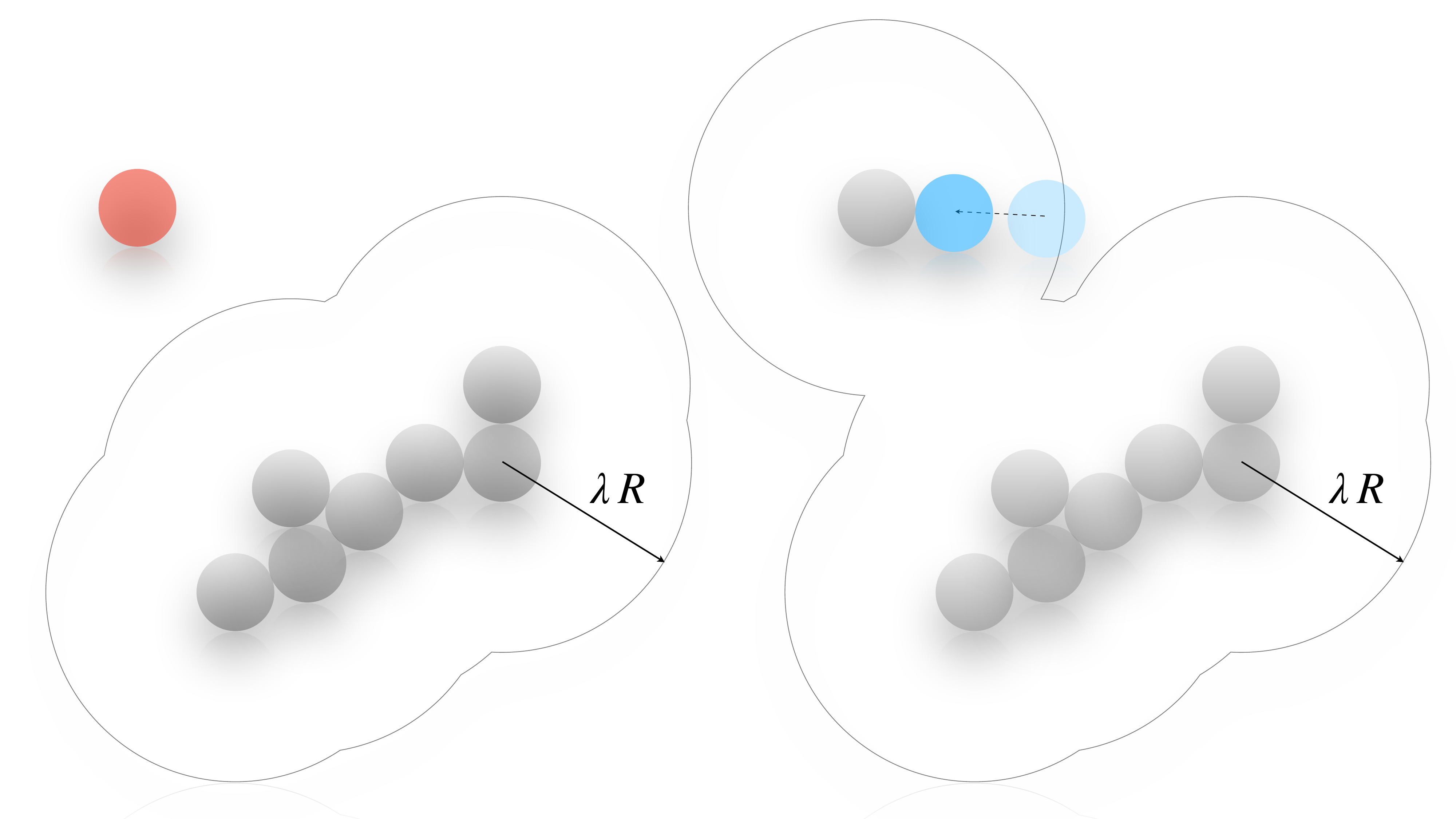}
    \caption{Aggregation algorithm: the red particle is not aggregated to the cluster of grey particles, but is nonetheless included in set $\mathcal{C}$: in this way, it can attract the blue particle and start growing its own cluster.}
    \label{fig:aggregation2}
\end{figure}

\section{Network resistivity}
\label{sec:ResistivityTemperatureCurves}
The first step is identifying the connected components of graph $\mathcal{G}$ that span the whole box (figure \ref{fig:Conductance}a). 

From each connected component containing terminal nodes (figure \ref{fig:Conductance}b), we build a node-admittance matrix $\mathbf{Y}_n$:
\begin{equation}
    \mathbf{Y}_n = \mathbf{B}\,\mathbf{Y}_b\,\mathbf{B}^T
\end{equation}
where $\mathbf{B}$ is the incidence matrix of the connected component and $\mathbf{Y}_b$ is the branch-admittance matrix \citep{chua1987linear}. We set a unitary conductance for each branch (edge of the graph) connecting two contact particles (figure \ref{fig:Conductance}b).

The \emph{node equation} is then
\begin{equation}
    \mathbf{Y}_n\, \mathbf{e} = \mathbf{i}_s
    \label{eq:node_equation}
\end{equation}
where $\mathbf{e}$ is the node voltage vector and $\mathbf{i}_s$ is the equivalent source vector of the currents.
\begin{figure}[htp]
    \centering
    \includegraphics[width=1\textwidth]{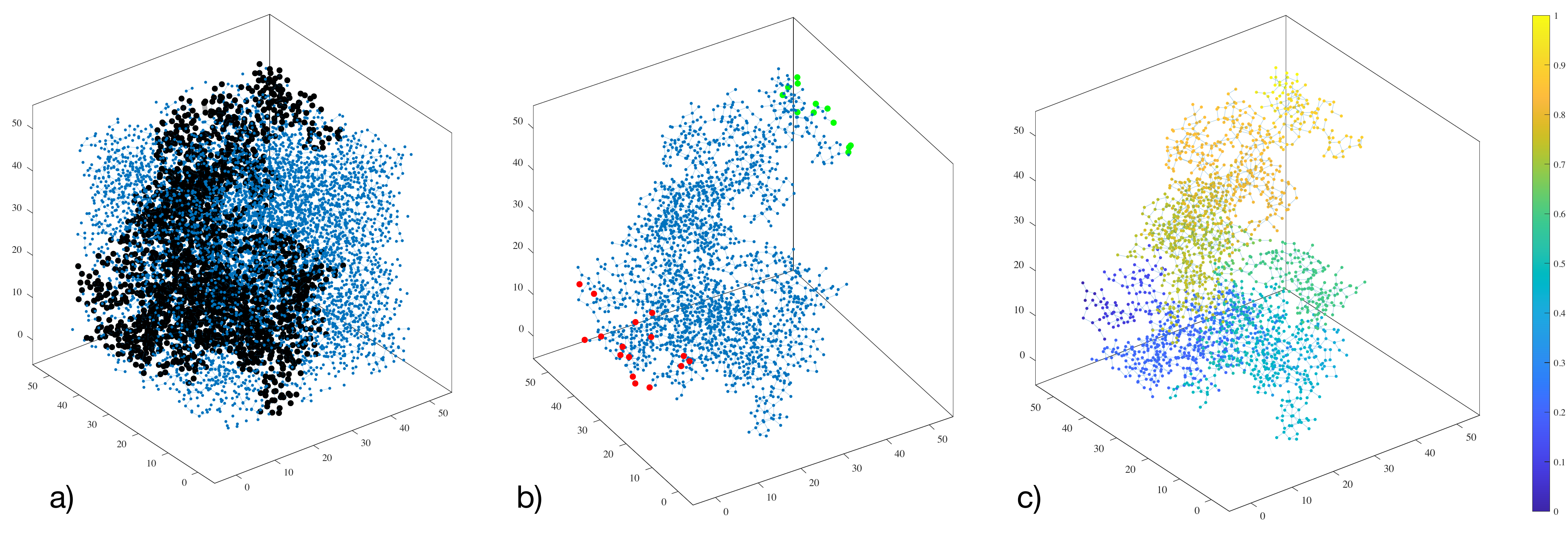}
    \caption{Effective resistivity of the network: a) network of contact particles, dots are the vertices (spheres' centers) and lines are the edges; black dots are a connected component with terminal nodes; b) the connected component in a) with terminal nodes (red and green dots), with voltage values $\mathbf{e}_e$, with red dots being the ground nodes (voltage set to zero) and green dots with voltage set to $1$; c) the resulting voltage distribution $\mathbf{e}$ at the network nodes.}
    \label{fig:Conductance}
\end{figure}
By definition, the connected components spanning the domain will have terminal nodes: we apply an unitary voltage difference between these nodes, for example $0$ on one side and $1$ on the opposite side (figure \ref{fig:Conductance}b). We indicate with subscript $_e$ such terminal nodes and with subscript $_i$ the remaining nodes, which we call \emph{internal} nodes.

Equation \eqref{eq:node_equation} can then be re-arranged in the following form
\begin{equation}
    \left( 
    \begin{array}{cc}
        \mathbf{G}_{ee} & \mathbf{G}_{ei} \\
        \mathbf{G}_{ei}^T & \mathbf{G}_{ii} 
    \end{array}
    \right)\,
     \left(
    \begin{array}{c}
         \mathbf{e}_e  \\
          \mathbf{e}_i
    \end{array}
    \right)
    =
    \left(
    \begin{array}{c}
         \mathbf{i}_e  \\
          \mathbf{0}
    \end{array}
    \right)
\end{equation}
where $\mathbf{e}_e$ is a known quantity and contains a voltage of $0$ if the node is on one side or a voltage of $1$ is the node is on the opposite side; $\mathbf{e}_i$ is unknown, and can be computed as
\begin{equation}
    \mathbf{e}_i = -{\mathbf{G}_{ii} }^{-1}\,\mathbf{G}_{ei}^T\, \mathbf{e}_e
\end{equation}
Once $\mathbf{e}_i $ are computed, we can obtain
\begin{equation}
    \mathbf{i}_e = \mathbf{G}_{ee}\, \mathbf{e}_e + \mathbf{G}_{ei}\,\mathbf{e}_i
\end{equation}
Figure \ref{fig:Conductance}c shows the resulting voltage $\mathbf{e}$.

The equivalent total current $I_eq$ is obtained by summing all the entries of vector $\mathbf{i}_e$ for nodes on one side: this sum will be equal to the sum of all the entries for the nodes on the opposite side, because of the Kirchhoff's current law.
The equivalent resistance $R_{eq} = \Delta V/I_{eq}$ and considering that we used $\Delta V=1$
\begin{equation}
    R_{eq} = \frac{1}{I_{eq}} \qquad I_{eq} =  \sum_{i:\mathbf{e}_{e_i}=1} \mathbf{i}_{e_i}
\end{equation}
We repeat this calculation for all the $n_{\mathrm{conn}}$ connected components having terminal nodes: the average of the equivalent resistances for these components is the effective resistivity of the network $\Bar{R}$
\begin{equation}
    \Bar{R} = \sum_{j=1}^{n_{\mathrm{conn}}} \frac{R_{eq_j}}{n_{\mathrm{conn}}}
\end{equation}

\section{Meshless algorithm}
\label{sec:GoverningEquations}
Consider a polymer matrix $\Omega$ and a set of $N$ non-overlapping spheres $\sph_I$ of radius $R_I \quad I=1,2,\ldots,N$. The polymer displacement vector is $\U(\x) = \left(\begin{array}{ccc}  u(\x) & v(\x) & w(\x)
\end{array}\right)^T$ with $\x \in \Omega$ and $\x = \left(\begin{array}{ccc}  x & y & z
\end{array}\right)^T$. The spheres are considered rigid, therefore having $3$ degrees of freedom, $\UU_I = \left( \begin{array}{ccc}
     U_I & V_I 
     & W_I
\end{array} \right)^T$ The weak formulation of linear elasticity reads:
\begin{align}
    &\int_\Omega \delta\T{\epsilon}^T\T{\sigma}\,\de \Omega - \int_\Omega \delta\T{\epsilon}^T\T{\sigma}_{TH}\,\de \Omega +\beta\,\int_{\Gamma_u} \delta\left(\U-\bar{\U}\right)^T\left(\U-\bar{\U}\right)\, \de \Gamma_u \nonumber \\
    &+ \alpha\,\sum_{I=1}^N \int_{\partial \sph_I} \delta\left(\U-\UU_I\right)^T\left(\U-\UU_I\right) \, \de \partial \sph_I = 0
    \label{eq:VariationalForm}
\end{align}
where $\delta(\cdot)$ indicates a variation, $\T{\epsilon}(\x)$ is the infinitesimal strain tensor in Voigt notation, $\T{\sigma}(\x)$ is the Cauchy stress tensor in Voigt notation, $\T{\sigma}_{TH}$ is the Cauchy stress tensor due to thermal strain $\T{\epsilon}_{TH}$, $\beta$ is a large penalty factor for the essential boundary conditions, $\U$ is the polymer displacement vector, $\bar{\U}$ is the vector of imposed displacements on the portion $\Gamma_u$ of the polymer boundary $\partial \Omega$, $\UU_I$ is the sphere's displacement vector, $\partial \sph_I$ is the surface of the $I-$th sphere and $\alpha$ is the bonding stiffness between polymer and sphere.

\begin{figure}[t]
    \centering
    \includegraphics[width=0.6\textwidth]{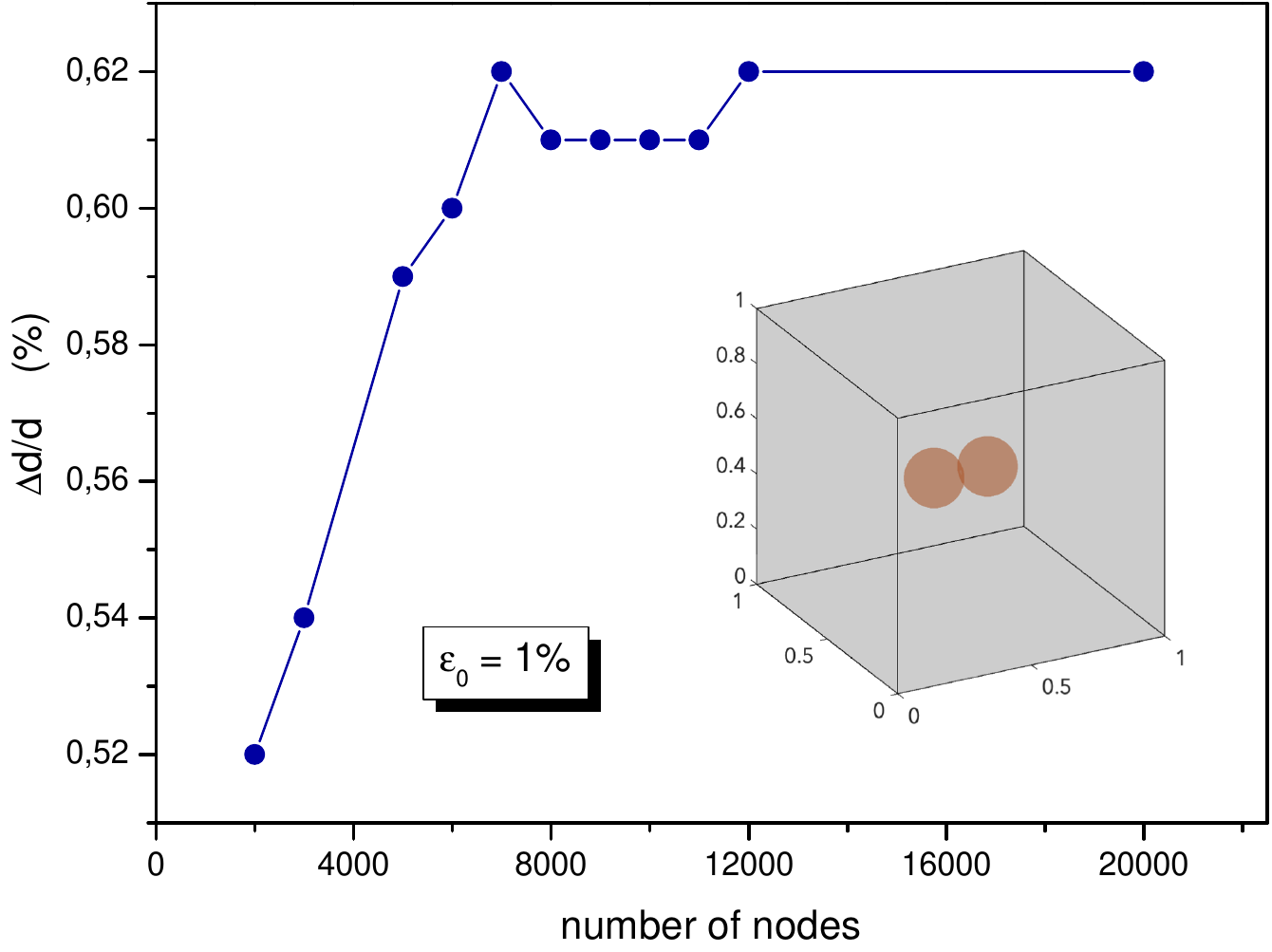}
    \caption{Relative variation of the distance between the centers of two spheres as a function of the number of nodes}
    \label{fig:convergence}
\end{figure}

We discretize $\Omega$ with $n$ balls $\disk$ of radius $\rho$. The centers of these balls are positioned at $\x_i$, such that $\Omega\subset \bigcup_{i=1}^n \disk$. We define compact support functions $\phi_i(\x)$ that are null outside $\disk$. Then, we use the \emph{ansatz} for the field variables
\begin{equation}
    u(\x) = \sum_{i=1}^n \phi_i(\x)\,u_i = \T{\phi}(\x)^T\,\left( \begin{array}{c}
        u_1 \\
        u_2 \\
        \ldots \\
        u_n 
    \end{array} \right) 
    \label{eq:uh}
\end{equation}
where $\phi_i(\x)$ are \emph{shape functions} and $u_i$ nodal coefficients. We use a meshless method (Reproducing Kernel Particle Method, RKPM \cite{Liu1995}) for the shape functions. The shape functions are \emph{corrected} weight functions $w$, where the values of $w$ is $1$ on $\x_i$ and zero outside a radius $\rho$, that is
\begin{equation}
w(\xi) = \begin{cases}
1-6\,\xi^2+8\, \xi^3-3\,\xi^4 &  \qquad 0\leq \xi \leq 1 \\ 
0 & \qquad \xi>1
\end{cases}
\label{eq:2kthspline} 
\end{equation}
where $\xi = \| \x - \x_i\|/\rho$. The shape functions are then
\begin{equation}
    \phi_i(\x) = C_i(\x)\,w\left(\frac{\x-\x_i}{\rho}\right)
\end{equation}
where $C_i(\x)$ are the \emph{corrections}
\begin{equation}
    C_i(\x) = \p\left(\mathbf{0}\right)^T\mathbf{M}(\x)^{-1}\,\p\left(\frac{\x-\x_i}{\rho}\right)
\end{equation}
where $\p(\x)$ is a vector of polynomial functions, for example, for linear polynomials
\begin{equation}
    \p\left(\x\right)^T = \left( \begin{array}{cccc}
        1 &  x & y & z 
    \end{array} \right) 
\end{equation}
and
\begin{equation}
    \mathbf{M}(\x) = \sum_{i=1}^n \p\left(\frac{\x-\x_i}{\rho}\right)\,\p\left(\frac{\x-\x_i}{\rho}\right)^T\,w\left(\frac{\x-\x_i}{\rho}\right)
\end{equation}

One of the advantages of meshless shape functions over finite element ones is that the approximation is constructed entirely over nodes without the need for a tessellation or a \emph{mesh}. This feature is crucial for domains with spheres in contact, where meshing between spheres is almost impossible. The price is a \emph{slight} increase in computational costs because a meshless method requires a neighbor search between nodes $\x_i$ and evaluation points $\x$.
Also, it requires a matrix inversion $\mathbf{M}(\x)^{-1}$ at each evaluation point. Nevertheless, kd-trees data structures minimize the neighbor search's computational burden \cite{Barbieri2012}. In addition, \cite{Barbieri2012} provides an iterative algorithm based on the Sherman-Morrison formula to invert the moment matrix quickly through explicit formulas.

Figure~\ref{fig:convergence} illustrates the performance of the meshless algorithm under varying the number of nodes. 
We consider two identical spheres of radius $0.1$ placed within a cubic box of unitary edge length (inset of Fig.~\ref{fig:convergence}. 
Initially, the two spheres are a contact, with a center-to-center distance of $d=0.2$. The main panel of 
Fig.~\ref{fig:convergence} displays the relative change ($\Delta d/d$) in the center-to-center distance as a function of the number 
of nodes for an applied thermal strain of $1$\%. The computations employ a bond stiffness of $100E$ between
the spheres and the matrix, where $E$ represents the Young modulus which we have set equal to $1$.
Notably, convergence is reached above $7000$ nodes. In the calculations detailed in the main text, $12000$ nodes were utilized.

In the following, we provide explicit expressions for the functions entering Eq.~\eqref{eq:VariationalForm}. 
Using \eqref{eq:uh}, the strain tensor is given by
\begin{equation}
    \T{\epsilon}(\x) = \left( \begin{array}{ccc}
        \dep{\T{\phi}}{x}^T & 0 & 0 \vspace{0.2cm} \\  
        0                  & \dep{\T{\phi}}{y}^T & 0 \vspace{0.2cm} \\
        0                  &   0                 & \dep{\T{\phi}}{z}^T \vspace{0.2cm} \\
        0 & \dep{\T{\phi}}{z}^T & \dep{\T{\phi}}{y}^T \vspace{0.2cm} \\
        \dep{\T{\phi}}{z}^T & 0 & \dep{\T{\phi}}{x}^T \vspace{0.2cm} \\
        \dep{\T{\phi}}{y}^T & \dep{\T{\phi}}{x}^T & 0
    \end{array} \right)\,\left( \begin{array}{c}
        u_1 \\
        u_2 \\
        \ldots \\
        u_n \\
        v_1 \\
        v_2 \\
        \ldots \\
        v_n \\
        w_1 \\
        w_2 \\
        \ldots \\
        w_n
    \end{array} \right) = \mathbf{B}(\x)\,\D
\end{equation}
We further assume a linear elastic isotropic constitutive model
\begin{equation}
    \T{\sigma} = \mathbf{C}\,\T{\epsilon} \quad \mathbf{C} = 
    \left( \begin{array}{cccccc}
       C_{11}  & C_{12} & C_{13} & 0 & 0 & 0 \\
       C_{12}  & C_{22} & C_{23} & 0 & 0 & 0 \\
       C_{13}  & C_{23} & C_{33} & 0 & 0 & 0 \\
       0       &   0    &   0    & C_{44} & 0 & 0 \\
       0       &   0    &   0    & 0 & C_{55} & 0 \\
       0       &   0    &   0    & 0 & 0 & C_{66} 
    \end{array}
    \right)
\end{equation}
and an isotropic thermal expansion $\T{\epsilon}_{TH} = \epsilon_{TH} \left( \begin{array}{cccccc}
    1 & 1 & 1 & 0 & 0 & 0 
\end{array} \right)^T$
Substituting in equation \eqref{eq:VariationalForm}, we obtain the following discretized equations of motion
\begin{equation}
   \left( \begin{array}{cc}
    \mathbf{K} + \beta\,\mathbf{V} + \alpha\,\mathbf{K}_{mm}   &  \alpha\,\mathbf{K}_{ms}\\
    \alpha\,\mathbf{K}_{ms}^T    & \alpha\,\mathbf{K}_{ss}
   \end{array}  \right)\,
   \left( \begin{array}{c}
      \mathbf{d}    \\
      \mathbf{D}
   \end{array} \right)= \left( \begin{array}{c}
\epsilon_{TH}\,      \mathbf{F}_{TH} + \beta\,\mathbf{F}_u   \\
      0
   \end{array} \right)
   \label{eq:DiscretizedEquationsOfMotion}
\end{equation}
with $\mathbf{D} = \left( \begin{array}{cccccccccccc}
    U_1 & U_2 & \ldots & U_N & V_1 & V_2 & \ldots & V_N & W_1 & W_2 & \ldots & W_N
\end{array} \right)^T$
where
\begin{equation}
    \mathbf{K} = \int_{\Omega} \mathbf{B}^T\,\mathbf{C}\,\mathbf{B}\,\de \Omega \quad \mathbf{V} = 
    \left(
    \begin{array}{ccc}
       \displaystyle \int_{\Gamma_u} \T{\phi}\,\T{\phi}^T \,\de \Gamma_u  &  & \\
         & \displaystyle \int_{\Gamma_u} \T{\phi}\,\T{\phi}^T \,\de \Gamma_u & \\
         &  & \displaystyle \int_{\Gamma_u} \T{\phi}\,\T{\phi}^T \,\de \Gamma_u
    \end{array}
    \right)
\end{equation}
\begin{equation}
    \mathbf{F}_{TH} = 
    \left(
    \begin{array}{c}
       \displaystyle \int_{\Omega} \left(C_{11} + C_{12} +  C_{13}\right)\,\dep{\T{\phi}}{x} \de\Omega   \vspace{0.3cm} \\
       \displaystyle \int_{\Omega} \left(C_{12} + C_{22} +  C_{23}\right)\,\dep{\T{\phi}}{y} \de\Omega   \vspace{0.3cm}  \\
       \displaystyle   \int_{\Omega} \left(C_{13} + C_{23} +  C_{33}\right)\,\dep{\T{\phi}}{z} \de\Omega   
    \end{array}
    \right) \qquad 
    \mathbf{F}_{u} =
    \left(
    \begin{array}{c}
       \displaystyle \int_{\Gamma_u} \T{\phi}\,(u-\bar{u}) \,\de \Gamma_u   \vspace{0.3cm} \\
       \displaystyle \int_{\Gamma_u} \T{\phi}\,(v-\bar{v}) \,\de \Gamma_u \vspace{0.3cm} \\
        \displaystyle \int_{\Gamma_u} \T{\phi}\,(w-\bar{w}) \,\de \Gamma_u
    \end{array}
    \right)
\end{equation}
\[
    \mathbf{K}_{mm} = 
    \left( 
    \begin{array}{ccc}
      \mathbf{K}_{mat,mat}   &  & \\
         &  \mathbf{K}_{mat,mat} & \\
         &  & \mathbf{K}_{mat,mat}
    \end{array}
    \right)
\]
\begin{equation}
    \mathbf{K}_{ms} = 
    \left( 
    \begin{array}{ccc}
      \mathbf{K}_{mat,sph}   &  & \\
         &  \mathbf{K}_{mat,sph} & \\
         &  & \mathbf{K}_{mat,sph}
    \end{array}
    \right) \quad
    \mathbf{K}_{ss} = 
    \left( 
    \begin{array}{ccc}
      \mathbf{K}_{sph,sph}   &  & \\
         &  \mathbf{K}_{sph,sph} & \\
         &  & \mathbf{K}_{sph,sph}
    \end{array}
    \right)
\end{equation}
\[
\mathbf{K}_{mat,mat} = \sum_{I=1}^N \int_{\sph_I}  \T{\phi}\,\T{\phi}^T \, \de \sph_I
\]
\begin{equation}     
     \mathbf{K}_{mat,sph} = \left( 
     \begin{array}{cccc}
      \int_{\sph_1} \T{\phi}\, \de \sph_1   &  \int_{\sph_2} \T{\phi}\, \de \sph_2   & \dots  &  \int_{\sph_N} \T{\phi}\, \de \sph_N    
     \end{array}
     \right) \quad
     \mathbf{K}_{sph,sph} = 
     \left( 
     \begin{array}{cccc}
       4\pi\,{R_1}^2   &  & & \\
          & 4\pi\,{R_2}^2 & & \\
          & & \ldots & \\
          & & & 4\pi\,{R_N}^2
     \end{array}
     \right)
\end{equation}

}

\end{document}